\documentstyle[12pt]{article}

\begin{document}

\vspace{1.5in}
\begin{center}
{{\LARGE \bf The Onset of Phase Transitions in Condensed Matter
and Relativistic QFT \footnote{To be published in Condensed Matter
Physics, 2000 } }}
\\
\baselineskip=18pt \vspace{1in} {\large R.J. Rivers, E.
Kavoussanaki,}
\\
\vspace{.1in} {{\it Blackett Laboratory, Imperial College, London
SW7 2BZ}}\\ \vspace{.1in}
 {\large G.\ Karra.}
\\
\vspace{.1in} {{\it Salomon Smith Barney International Limited,
London SW1W OSB}}\\
\end{center}

\vfill

\begin{abstract}
Kibble and Zurek have provided a unifying causal picture for the
appearance of topological defects like cosmic strings or vortices
at the onset of phase transitions in relativistic QFT and
condensed matter systems respectively. There is no direct
experimental evidence in QFT, but in condensed matter the
predictions are largely, but not wholly, supported in superfluid
experiments on liquid helium. We provide an alternative picture
for the initial appearance of strings/vortices that is
commensurate with all the experimental evidence from condensed
matter and consider some of its implications for QFT.


\end{abstract}

\section{The Problem}

One of the great successes in quantum field theory (QFT) in the
last two decades has been the realisation of the unification of
the disparate forces of quantum physics through spontaneous
symmetry breaking.  Even without a direct observation of the Higgs
meson for electroweak unification, to date, supplementary evidence
in support of the Standard Model of Glashow, Salam and Weinberg
suggests that it is only a matter of time before Higgs particles
will be found.

Analogy with condensed matter physics would suggest that this and
other symmetries were not always broken but that, in the very
early, and very hot universe, they were restored.  With several
levels of unification anticipated we expect several changes of
phase to have occurred, sequential partial orderings of an
initially disordered state. Although the effects of most of the
transitions can only be inferred indirectly, the last change of
phase relevant to particle physics, the hadronisation of the
quark-gluon plasma, is accessible at the heavy-ion collider
facility at Brookhaven (RHIC), just about to begin to produce
data. Although this transition is too difficult to address in this
article, it provides a huge incentive to understand the way in
which phase transitions occur in QFT.

Consider a situation in which the  symmetry group  of the theory
is broken, on cooling through the critical temperature $T_c$,
 by the degenerate groundstate manifold of the order-parameter fields.
Using the tools of {\it equilibrium} thermal field theory (TFT) we
can determine the nature of the transition. At late times after
the transition the fields are ordered on large scales, in that
they adopt a single value from this degenerate set over a large
spatial region.

To understand how this is achieved requires that we go beyond
equilibrium TFT. In practice, we often know remarkably little
about the dynamics of thermal systems. In particular, we shall
restrict our discussion to the {\it onset} of such phase
transitions, the very early times after the implementation of  a
transition when the scalar fields are only just beginning to
become ordered.

A simple question to ask is the following: In principle, the field
correlation length diverges at a continuous transition.  In
practice, it does not.  What happens?  This is relevant for
transitions that leave topological defects like walls, monopoles,
vortices, or textures in their wake since we might expect
'defects' to be just that, entities whose separation is
characterised by the correlation length.  If this were simply so,
an understanding of correlation lengths would lead to a prediction
for defect densities.  Conversely, a measurement of defect
densities would be a measurement of correlation lengths. Vortices,
in particular (cosmic strings), can be important for structure
formation in the universe. Because of their implications for
astrophysics, estimates of early field ordering in  their presence
have been made by Kibble\cite{kibble1,kibble2}, using simple
causal arguments.

There are great difficulties in converting such predictions for
the early universe into experimental observations. Zurek
suggested\cite{zurek1} that similar arguments were applicable to
condensed matter systems for which direct experiments on defect
densities, in particular vortex densities, could be performed.
This has lead to a burst of activity from condensed matter
experimentalists developing experiments to test these predictions.
To date, agreement with the Zurek predictions for a variety of
condensed matter systems is good, but not total.

The problems are twofold, in that we need a better understanding
of the predictions and of the experiments. In this article we
shall only consider the first. Many of our observations have been
discussed in more detail elsewhere, and our aim here is to give an
overview. However, there is so much work in progress in this area
that we do not claim to be exhaustive.

\section{The Predictions}

The predictions of Kibble and Zurek for the way in which the
fields freeze in at the onset of a transition are simply expressed
in terms of causality.  As such, they are generic. However, to
quantify them it is convenient to consider simple field theories
in which they can be realised.

In principle, the simplest transitions are temperature quenches.
In practice, the circumstances in which transitions occur are
poorly understood. In the early universe we do not necessarily
begin from a thermal distribution. More awkwardly, in condensed
matter physics (and heavy-ion collisions) we do not have a
spatially uniform temperature. However, for the sake of argument
we assume a uniform temperature, and leave more realistic
scenarios until later.

There is a further problem. In condensed matter systems the nature
of the fields in the Landau-Ginzburg effective theory is
prescribed, and their number is small. In the early universe we
have little constraint on our model making beyond a feeling that,
despite the ubiquity of gauge fields at later times, at early
times scalar fields are important.

\subsection{The models}

As we shall see later, most experiments involve vortices and, with
this in mind, we adopt the simplest theory, that of a {\it single}
complex scalar field $\phi$ in three spatial dimensions with a
wine-bottle potential. The transition is continuous\footnote{There
is no evidence for strong first order, or discontinuous,
transitions in QFT. The mechanism for such transitions, bubble
nucleation, would lead to a very different onset than that
considered here. }.

That is, we assume that the qualitative dynamics are conditioned
by the field's {\it equilibrium} free energy, of the form
\begin{equation}
F(T) = \int d^{3}x\,\,\bigg(|\nabla\phi |^{2} + m^{2}(T)|\phi
|^{2} + \lambda |\phi |^{4}\bigg). \label{FR}
\end{equation}
Intentionally, with the early universe in mind we have written
$F(T)$ of Eq.\ref{FR} in the form appropriate for a relativistic
quantum field. The coefficient $m^{2}(T)$ has the interactions
with the real particles of the heat bath taken into account, and
vanishes at $T=T_c$. We are considering the $\phi$ field as an
open system, in which changes in the external environment lead to
changes in the parameter $m^{2}(T)$, which takes the value $-M^2$
when $T=0$. For relativistic QFT the change in temperature that
leads to the change in the sign of $m^{2}$ is most simply
understood as a consequence of the system expanding. Thus, in the
early universe, once thermalisation is possible, a weakly
interacting relativistic plasma at temperature $T\gg M$ has an
entropy density $s\propto T^{3}$. As long as thermal equilibrium
can be maintained, constant entropy $S$ per comoving volume,
$S\propto s a(t)^{3}$, gives $T\propto a(t)^{-1}$ and falling, for
increasing scale factor $a(t)$.

Specifically, at time $t$ the temperature $T(t)$ satisfies
\begin{equation}
tT(t)^2 = m_{P}\,\bigg(\frac{45}{16\pi^{3}N^{*}}\bigg)^{1/2}
\label{temp}
\end{equation}
where $m_{P}$ is the Planck mass, and $N^*$ is the number of
effective field degrees of freedom.  Models that attempt to take
inflation into account, however, lead to 'preheating' that is not
Boltzmannian\cite{juan}.  Nonetheless, even in such cases it is
possible to isolate an effective temperature for long-wavelength
modes.  This is all that is necessary, but is too sophisticated
for the simple scenarios that we shall present here. Even the
inclusion of an FRW metric would complicate the issue at this
stage, and we assume flat space-time, with decreasing temperature.
Prior to the transition, we assume a uniform temperature
$T>T_{c}$, for which $m^{2} (T) > 0$. After the transition,
$m^{2}(0) = - M^{2}< 0$ enforces the $U(1)$ symmetry-breaking,
with field expectation values $\langle\phi\rangle = \pm\eta$,
$\eta^{2} = M^{2}/\lambda$. The Compton wavelength $\xi_0 =
M^{-1}$ is the natural distance scale. For the sake of argument we
assume mean-field behaviour $m^{2}(T) = M^{2}(T/T_{c} -1)$,
whereby the equilibrium correlation length $\xi_{eq}(T) =
|m(T)|^{-1} = \xi_{0}(T/T_{c} -1)^{-1/2}$.

Equally well, after rescaling, $F$ could be the Ginzburg-Landau
free energy
\begin{equation}
F(T) = \int d^{3}x\,\,\bigg(\frac{\hbar^{2}}{2m}|\nabla\phi |^{2}
+\alpha (T)|\phi |^{2} + \beta |\phi |^{4}\bigg) \label{FNR}
\end{equation}
for a non-relativistic condensed matter field, in which the
chemical potential $\alpha (T) = \alpha_{0}(T/T_{c}-1)$ vanishes
again at the critical temperature $T_{c}$. In this case we
envisage the change in $\alpha (T)$ as a consequence of an
external  cooling of the system or a change in the pressure of the
system that leads to a change in $T_c$. In either case, we again
assume circumstances in which, in a finite time, $T/T_c$ varies
from greater than unity to less.  The fundamental length scale
$\xi_{0}$ is given from  Eq.\ref{FNR} as
 $\xi_{0}^{2} = \hbar^{2}/2m\alpha_{0}$ whereby
 $\xi_{eq}(T) = \xi_{0}(T/T_{c} -1)^{-1/2}$
as before.  The Gross-Pitaevski theory\cite{zurek1} suggests a
natural time-scale $\tau_{0} = \hbar /\alpha_{0}$. When, later, we
adopt the time-dependent Landau-Ginzburg (TDLG) theory we find
this still to be true, empirically, at order-of-magnitude level,
and we keep it.

The minima of the final potential  now constitute the circle $\phi
= \eta e^{i\alpha}$, where $\eta^{2} = M^{2}/\lambda$ for
Eq.\ref{FR}, and $\eta^{2} = \alpha_{0}/\beta$ for Eq.\ref{FNR}.
When the transition begins $\phi$ begins to fall into the valley
of the potential, choosing a random phase. Although this randomly
chosen phase will vary from point to point we expect domains
across which the phase is roughly constant. How this collapse
takes place determines the size of the first identifiable domains.
It was suggested by Kibble and Zurek that this size is essentially
the equilibrium field correlation length $\xi_{eq}$ at some
appropriate temperature close to the transition. Two very
different mechanisms have been proposed for estimating this size.

\subsection{A first guess: Thermal activation}

In early work on transitions it was assumed\cite{kibble1}  that
initial domain size was fixed in the Ginzburg regime, identifiable
from {\it equilibrium} theory.  By this we mean the following.
Suppose the temperature $T(t)$ varies sufficiently slowly with
time $t$ that it makes sense to replace $V(\phi ,T)$ by $V(\phi ,
T(t))$.  Well away from the transition this is justified, but
close to the transition it is not. Once we are below $T_{c}$, and
the central hump in $V(\phi , T(t))$ is forming, the Ginzburg
temperature $T_{G} <T_{c}$ signals the temperature above which
there is a significant probability for thermal fluctuations over
the central hump on the scale of the correlation length at that
temperature.  Most simply, it is determined by the condition
\begin{equation}
\Delta V(T_{G})\xi_{eq}^{3}(T_{G})\approx k_{B}T_{G}
\end{equation}
where $\Delta V(T)$ is the difference between the central maximum
and the minima of $V(\phi ,T)$. That is,
\begin{equation}
\xi_{eq}(T_{G}) = O\bigg(\frac{\xi_{0}}{(1 -
T/T_{c})^{1/2}}\bigg), \label{xiG}
\end{equation}

Whereas, above $T_{G}$ there will be a population of 'domains',
fluctuating in and out of existence, at temperatures below $
T_{G}$ fluctuations from one minimum to the other become
increasingly unlikely. For the relativistic theory of Eq.\ref{FR},
in units in which $k_{B} $ is unity, we find $|1-T_{G}/T_{c}| =
O(\lambda )$, whereby
\begin{equation}
\xi_{eq}(T_{G}) = O\bigg(\frac{1}{\lambda T_c}\bigg) \label{tom1}
\end{equation}
On the other hand, for non-relativistic condensed matter, we find
\begin{equation}
1-T_{G}/T_{c} = (\beta k_{B}T_{c}/\alpha_{0}^{2}\xi_{0}^{3})^{2}
\end{equation}

It was originally suggested by Kibble\cite{kibble1} that we
identify $\xi_{eq}(T_{G})$ with the scale at which stable domains
begin to form. We shall show this to be incorrect, for quenches
that are not too slow.
However, we will find that strong thermal fluctuations do have a
role, particularly in $^{4}He$, for which the whole experiments
take place within the Ginzburg regime. This is an issue that
requires more than equilibrium physics. The most simple dynamical
arguments invoke causality.

\subsection{Causality in QFT and condensed matter}

The first application of causality was again due to
Kibble\cite{kibble3}. Whatever the details of the microscopic
physics, causality sets an upper limit over which the field can be
correlated in the causal horizon with diameter $r(t)\approx 2t$.
In the vicinity of the transition, whether at $T=T_G$ or not, this
gives a correlation length $\xi (t) \leq r(t)$ or, from
Eq.\ref{temp}
\begin{equation}
\xi\leq \bigg(\frac{45}{4\pi^3 N^*}\bigg)^{1/2}\frac{m_P}{T_c^{2}}
\label{tom2}
\end{equation}
where $N^*$ the number of field degrees of freedom at $T_c$.  This
was used to bound monopole density in the early universe. Although
the lack of correlation in the field phase in different causal
horizons demonstrates the existence of a field structure that will
naturally lead to defects, beyond that it is not a helpful guide.

If this invocation of causality, or the Ginzburg criteria, attempt
to set scales once the critical temperature has been {\it passed},
other causal arguments attempt to set scales {\it before} it is
reached. Again suppose that $T/T_{c}$ varies in time as a
consequence of the change in the environment. We have seen that
$\xi_{eq}(T(t))$, obtained by inserting the time-dependence of
$T/T_{c}$ explicitly,  diverges at $T(t) =T_{c}$, which we suppose
happens at $t=0$. This cannot be the case for the true correlation
length $\xi (t)$, which can only grow so far in a finite time.
Initially, for $t<0$, when we are far from the transition, we
again assume effective equilibrium, and the field correlation
length $\xi (t)$ tracks $\xi_{eq}(T(t))$ approximately. However,
as we get closer to the transition, $\xi_{eq}(T(t))$ begins to
increase arbitrarily fast and the adiabatic approximation breaks
down. This is the {\it largest} value that $\xi_{eq}$ attains.
This {\it largest} value {\it prior} to the transition corresponds
to the {\it smallest} value of the field correlation length {\it
after} the transition. We shall argue later that the argument is
too simple but, nonetheless it is a plausible starting point. The
question then becomes: how large does the field correlation get in
practice? Whether in QFT or condensed matter physics there is a
maximum speed at which $\xi (t)$  can grow on purely causal
grounds. As a crude upper bound, in QFT the true correlation
length $\xi (t)$ fails to keep up with $\xi_{eq}(T(t))$ by the
time $-{\bar t}$ at which $\xi_{eq}$ is growing at c=1, the speed
of light, $d\xi_{eq}(T(-{\bar t}))/dt =1$.  It was suggested,
again by Kibble\cite{kibble2}, that once we have reached this time
$\xi (t)$ {\it freezes} in, remaining approximately constant until
the time $t\approx +{\bar t}$ after the transition when it once
again becomes comparable to the now {\it decreasing} value of
$\xi_{eq}$. The correlation length ${\bar\xi} =\xi_{eq}({\bar
t})=\xi_{eq}(-{\bar t})$ is argued to provide the scale for the
minimum domain size {\it after} the transition.

Specifically, if we assume a time-dependence $m^{2}(t) =
-M^{2}t/t_Q$ in the vicinity of $t=0$, when the transition begins
to be effected, then the causality condition gives
$t_C=t_{Q}^{1/3}(2M)^{-2/3}$.  As a result,
\begin{equation}
M\xi_{eq}({\bar t}) = (M\tau_{0})^{1/3}, \label{xiC0}
\end{equation}
which we write as
\begin{equation}
{\bar\xi} = \xi_{eq}({\bar t}) =
\xi_{0}\bigg(\frac{\tau_{Q}}{\tau_{0}}\bigg)^{1/3} \label{xiC}
\end{equation}
where $\tau_{0}  = \xi_{0}= M^{-1}$ are the natural time and
distance scales.  In contrast to Eq.\ref{xiG}, Eq.\ref{xiC}
depends explicitly on the quench rate, as we would expect.  For
$\tau_{Q}\gg \tau_{0}$ the field is correlated on a scale of many
Compton wavelengths.

This approach of Kibble was one of the motivations for a similar
analysis by Zurek\cite{zurek1} of transitions with scalar order
parameters in condensed matter. Explicitly, in the Ginzburg-Landau
free energy Eq.\ref{FNR}, $\alpha (T)$ also vanishes at the
critical temperature ${T_c}$.   The only difference is that, in
the causal argument, the speed of light should be replaced by the
speed of (second) sound, with different critical index.

Explicitly, let us again assume the mean-field result $\alpha (T)
= \alpha_{0}\epsilon (T)$, where $\epsilon = (T/T_c -1)$, remains
valid as $T/T_c$ varies with time $t$ as $\alpha (t)=\alpha
(T(t))=-\alpha_{0} t/\tau_{Q}$ in the vicinity of $T_c$.   It
follows that the equilibrium correlation length $\xi_{eq} (t)$ and
the relaxation time $\tau (t)$ diverge when $t$ vanishes as
\begin{equation}
\xi_{eq} (t) = \xi_{0}\bigg|\frac{t}{\tau_{Q}}\bigg|^{-1/2},
\,\,\tau (t) = \tau_{0}\bigg|\frac{t}{\tau_{Q}}\bigg|^{-1},
\end{equation}
in terms of $\xi_0$ and $\tau_0$ as given earlier.

 The speed of
sound is $c(t) =\xi_{eq} (t)/\tau (t)$, slowing down as we
approach the transition as $|t|^{1/2}$. The causal counterpart to
$d\xi_{eq} (t)/dt = 1$ for the relativistic field is $d\xi_{eq}
(t)/dt = c(t)$.  This is satisfied at $t=-{\bar t}$, where ${\bar
t} =\sqrt{\tau_{Q}\tau_{0}}$, with corresponding correlation
length
\begin{equation}
{\bar\xi} =\xi_{eq}({\bar t}) =\xi_{eq}(-{\bar t}) =
\xi_{0}\bigg(\frac{\tau_{Q}}{\tau_{0}}\bigg)^{1/4}. \label{xiZ}
\end{equation}
(cf. Eq.\ref{xiC}). A variant of this argument\cite{zurek1} that
gives essentially the same results is obtained by comparing the
quench rate directly to the relaxation rate of the field
fluctuations.  We stress that, yet again, the assumption is that
the length scale that determines the initial correlation length of
the field freezes in {\it before} the transition begins. Whatever,
the field is already correlated on a scale of many Compton
wavelengths when it begins to unfreeze.

\subsection{Experimental predictions}

The end result of the simple causality arguments is that, both for
QFT and condensed matter, when the field begins to order itself
its correlation length has the form
\begin{equation}
{\bar\xi} = \xi_{0}\bigg(\frac{\tau_{Q}}{\tau_{0}}\bigg)^{\gamma}.
\label{xiKZ}
\end{equation}
for appropriate $\gamma$.  In fact, the powers of Eq.\ref{xiC} and
Eq.\ref{xiZ} are mean-field results, changed on implementing the
renormalisation group. Whereas, for $^3 He$, the critical
behaviour of Eq.\ref{xiZ} survives, for $^4 He$, $\gamma =1/3$. On
the other hand, for QFT the relevant parameter is
${\bar\xi}T_{c}$, the ratio of the maximum correlation length to
the thermal length $\beta_{c} = T_{c}^{-1}$. In equilibrium
theory, when the temperature is high enough the theory is
essentially three-dimensional, with critical index $\gamma$
different from its {\it four}-dimensional mean-field value.
 At our level of discussion, it is sufficient to keep ({\it
four}-dimensional) mean-field values.

As we said earlier, when the transition begins $\phi$ begins to
fall into the valley of the potential, choosing a random phase.
This randomly chosen phase will vary from point to point leading
to approximate  domains across which
 the phase is roughly constant.
Such domains will meet at defects, in this case vortices, tubes of
'false' vacuum $\phi\approx 0$, around which the field phase
changes by $\pm 2\pi$. In an early universe context these are
'cosmic strings', but we shall not consider their properties here.

Correlation lengths in the early universe are not amenable to
direct observation.  Kibble made a second {\it assumption}, that
the correlation
 length Eq.\ref{xiC}
 {\it also} sets the scale for the typical minimum intervortex
distance at the time that vortices are produced.

That is, the {\it initial} vortex density $n_{def}$ is
\begin{equation}
n_{def} = O\bigg(\frac{1}{{\bar\xi}^{2}}\bigg)=
\frac{1}{f^{2}\xi_{0}^{2}}\bigg(\frac{\tau_{0}}{\tau_{Q}}\bigg)^{2\gamma},
\label{ndef}
\end{equation}
for $\gamma = 1/3$, where $f = O(1)$ estimates the fraction of
defects per 'domain'. Equivalently, the length of vortices in a
box volume $v$ is $O(n_{def}v)$.

What is striking and suspect about these predictions is that they
are universal.  They do not use any information about the strength
($\lambda$ or $\beta$) of the interactions, and hence the
magnitude of the order parameter after the transition or, in
consequence, the existence of the Ginzburg regime.

Since $\xi_{0}$ also measures cold vortex thickness, $\tau_{Q}\gg
\tau_{0}$ corresponds to a measurably large number of widely
separated vortices. For the early universe Kibble deduced
\begin{equation}
{\bar\xi} =
\bigg(\frac{m_{P}}{\sqrt{N^*}M^{2}T_{c}^{2}}\bigg)^{1/3},
\label{tom3}
\end{equation}
where, as before, $m_P$ is the Planck mass, and $N^*$ the number
of field degrees of freedom at $T_c$\footnote{We note that,
despite their different origins, the predictions
\ref{tom1},\ref{tom2},\ref{tom3} may not differ so hugely from one
another for very small coupling.}.
 If it could be argued
that this initial network behaves classically then, thereafter,
the density will reduce due to the collapse of small loops,
intersections chopping off loops which in turn collapse, and
vortex straightening so as to reduce the gradient energy of the
field.

However, even if cosmic strings were produced in so simple a way
in the very early universe it is still not possible to compare the
density Eq.\ref{ndef} with experiment. What is more amenable to
experiment, in principle, is the length distributions of string
networks, and their ability to show scaling behaviour.  This only
impinges indirectly on the correlation length of the field, but
would have to be commensurate with any density calculations.

It was Zurek\cite{zurek1} who first suggested that, if this
relationship Eq.\ref{ndef} between defect density and correlation
length were true, it could be tested directly in condensed matter
systems, particularly in liquid helium.

\section{Experiments in Condensed Matter}

\subsection{Vortices in superfluid helium}

Vortex lines in both superfluid $^{4}He$ and $^{3}He$ are
analogues of global cosmic strings. A crude but effective model is
to treat the system as composed of two fluids, the normal fluid
and the superfluid, which has zero viscosity. In $^{4}He$ the bose
superfluid is characterised by a complex field $\phi$, whose
squared modulus $|\phi |^{2}$ is the superfluid density.  The
superfluid fraction is unity at absolute zero, falling to zero as
the temperature rises to the lambda point at 2.17K.  The
Landau-Ginzburg theory for $^{4}He$ has, as its free energy $F(T)$
of Eq.\ref{FNR}.

The situation is more complicated, but more interesting, for
$^{3}He$, which becomes superfluid at the much lower temperature
of 2 mK. The reason is that the $^{3}He$ is a {\it fermion}.  Thus
the mechanism for superfluidity is very different from that of
$^{4}He$.  Somewhat as in a BCS superconductor, these fermions
form the counterpart to Cooper pairs.  However, whereas the
(electron) Cooper pairs in a superconductor form a $^{1}S$ state,
the $^{3}He$ pairs form a $^{3}P$ state. The order parameter
$A_{\alpha i}$ is a complex $3\times 3$ matrix $A_{\alpha i}$.
There are two distinct superfluid phases, depending on how the
$SO(3)\times SO(3)\times U(1)$ symmetry is broken. If the normal
fluid is cooled at low pressures, it makes a transition to the
$^{3}He-B$ phase, in which $A_{\alpha i}$ takes the form
$A_{\alpha i} = R_{\alpha i}(\omega)e^{i\Phi}$, where $R$ is a
real rotation matrix, corresponding to a rotation through an
arbitrary $\omega$\cite{volovik2} \footnote{At large distance
scales there is a complication in that the small spin-orbit
coupling becomes important, to fix $\omega$ at $arc cos (-1/4)$,
but this will not concern us here.}.

The Landau-Ginzburg free energy is, necessarily, more
complicated\cite{Bunkov}, but the effective potential $V(A_{\alpha
i}, T)$ has the diagonal form $V(A, T) = \alpha (T)|A_{ai}|^{2} +
O(A^{4})$ for small fluctuations, and this is all that we need for
the production of vortices at very early times. Beyond that it can
be mimicked by Eq.\ref{FNR} for our purposes (e.g. see
\cite{kopnin}). Thus the Zurek analysis leads to the prediction
Eq.\ref{ndef}, as before, for appropriate $\gamma$. However, for
$^{3}He$ the mean-field approximation is good and the mean-field
critical index $\gamma = 1/4$ is not renormalised, whereas for
$^{4}He$ a better value is $\gamma = 1/3$, as for the naive
relativistic theory.

\subsection{Experiments in $^{3}He$.}

Although $^{3}He$ is more complicated to work with, the
experiments to check Eq.\ref{ndef} are cleaner in that, because
the nucleus has spin $1/2$, even individual vortices can be
detected by magnetic resonance. Further, because vortex width is
many atomic spacings the Landau-Ginzburg theory is reliable.

So far, experiments have been of two types.  In the Helsinki
experiment\cite{helsinki} superfluid $^{3}He -B$ in a rotating
cryostat is bombarded by slow neutrons.  Each neutron entering the
chamber releases 760 keV, via the reaction $n + ^{3}He\rightarrow
p + ^{3}He + 760 keV$.  The energy goes into the kinetic energy of
the proton and triton, and is dissipated by ionisation, heating a
region of the sample above its transition temperature.  The heated
region then cools back through the transition temperature,
creating vortices. Vortices above a critical size (dependent on
the angular velocity of the cryostat) grow and migrate to the
centre of the apparatus, where they are counted by an NMR
absorption measurement.  Suffice to say that the quench is very
fast, with $\tau_{Q}/\tau_{0} = O(10^{3})$.   Agreement with
Eq.\ref{ndef} and Eq.\ref{xiZ} is very good, at the level of less
than an order of magnitude. This is even though it is now
argued\cite{kopnin} that the Helsinki experiment should {\it not}
show agreement because of the geometry of the heating event.

The second type of experiment has been performed at Grenoble and
Lancaster\cite{grenoble}.  Rather than count individual vortices,
the experiment detects the total energy going into vortex
formation. As before, $^{3}He$ is irradiated by neutrons.  After
each absorption the energy released in the form of quasiparticles
is measured, and found to be less than the total 760 keV. This
missing energy is assumed to have been expended on vortex
production.  Again, agreement with Zurek's prediction
Eq.\ref{ndef} and Eq.\ref{xiZ} is good.

\subsection{Experiments in $^{4}He$.}

The experiments in $^{4}He$, conducted at Lancaster, follow
Zurek's original suggestion.  The idea is to expand a sample of
normal fluid helium, in a container with bellows, so that it
becomes superfluid at essentially constant temperature. That is,
we change $1-T/T_c$ from negative to positive by reducing the
pressure, thereby increasing $T_c$. As the system goes into the
superfluid phase a tangle of vortices is formed, because of the
random distribution of field phases.  The vortices are detected by
measuring the attenuation of second sound within the bellows.
Second sound scatters off vortices, and its attenuation gives a
good measure of vortex density.  A mechanical quench is slow, with
$\tau_{Q}$ some tens of milliseconds, and $\tau_{Q}/\tau_{0} =
O(10^{10})$.  Two experiments have been
performed\cite{lancaster,lancaster2}.  In the first fair agreement
was found with the prediction Eq.\ref{ndef}, although it was not
possible to vary $\tau_{Q}$.  However, there were potential
problems with hydrodynamic effects at the bellows, and at the
capillary with which the bellows were filled.  A second
experiment, designed to minimise these and other  problems has
failed to see any vortices whatsoever.

Some care is needed. Not only is the Landau-Ginzburg effective
theory more suspect for $^{4}He$, but its Ginzburg regime is so
wide, at $O(1K)$ that the transition takes place entirely within
it.

\subsection{Quenching in an annulus}

Experiments for vortex densities are problematical in that a
closely bound vortex-antivortex pair gives a count of 2 right up
to annihilation, whereas their {\it topological} charge remains
zero. It is possible to devise experiments that count topological
charge. Consider a closed path in the bulk superfluid with
circumference $C\gg \xi (t)$. Naively, the number of 'regions'
through which this path passes in which the phase is correlated is
${\cal N} = O(C/\xi (t))$. Assuming an independent choice of phase
in each 'region', the r.m.s phase difference along the path is
\begin{equation}
\Delta\theta_{C} \approx\sqrt{{\cal N}} = O(\sqrt{C/{\bar\xi}}).
\label{rphase}
\end{equation}

If we now consider a quench in an annular container of similar
circumference $C$ of superfluid $^{4}He$ and radius $l\ll C$,
Zurek suggested\cite{zurek1} that the phase locked in is {\it
also} given by Eq.\ref{rphase}, with ${\bar \xi}$ of Eq.\ref{xiZ}.
Since the phase gradient is directly proportional to the superflow
velocity we expect a flow after the quench with r.m.s velocity
\begin{equation}
\Delta v =
O\bigg(\frac{\hbar}{m}\sqrt{\frac{1}{C{\bar\xi}}}\bigg).
\label{v1}
\end{equation}
provided $l = O({\bar\xi})$.  Although in bulk fluid this
superflow will disperse, if it is constrained to a narrow annulus
it should persist, and although not large is measurable, in
principle. Specifically, in the units of Zurek\cite{zurek1}
\begin{equation}
\Delta v = O((\tau_{Q}[\mu s])^{-\nu /4}/\sqrt{C[cm]}) \label{vf0}
\end{equation}
where $\tau_{Q}$ is, typically, tens of milliseconds and $C$ in
centimetres. $\nu = 1/2$ is the mean-field critical exponent
above.  In principle $\nu$ should be renormalised to $\nu = 2/3$,
but the difference to $\Delta v$ is negligable.  With such a small
index the result is almost independent of quench rate.

In practice there are difficulties in performing annular
measurements in $^{4}He$. A similar, but easier, experiment can be
performed on annular Type-II superconductors, on cooling through
their critical temperatures. The relevant free energy
is\cite{zurek1} the extension of $F$ of Eq.\ref{FNR},
\begin{equation}
F(T) = \int d^{3}x\,\,\bigg(\frac{1}{4m}|-i\hbar\nabla\phi
-\frac{2e}{c}{\bf A}|^{2} +\alpha (T) |\phi |^{2} +
\frac{1}{4}\beta |\phi |^{4}\bigg) + \frac{B^{2}}{8\pi}.
\label{SC}
\end{equation}
{\bf A} is the vector potential in the Coulomb gauge, and ${\bf B}
= \nabla\wedge{\bf A}$.

If we can initially ignore the effects of the gauge field as a
temperature quench is imposed through a change in $\alpha (T)$, as
before, the result Eq.\ref{rphase} persists. For perimeter $C$ the
variance in the number of flux quanta produced spontaneously is
\begin{equation}
\Delta
N_{C}=\frac{1}{2\pi}\Delta\theta_{C}\approx\frac{1}{2\pi}\sqrt{\frac{C}{{\bar\xi}}}.
\label{rphase2}
\end{equation}

In fact, it is more convenient to quench annular Josephson
Junctions, in which two identical rings of  superconductor are
held apart by an oxide layer through which Cooper pairs can
tunnel. If $\theta_{1}$ and $\theta_{2}$ are the phases of $\phi$
on the upper and lower rings then, once the transition has taken
place the tunneling current has the form
\begin{equation}
J = J_{c} sin (\theta_{1}-\theta_{2})
\end{equation}
and the theory is described by a dissipative Sine-Gordon equation.

The kinks of this equation are the 'fluxons' of the Josephson
Junction and are easy to observe experimentally\cite{monaco}. The
variance in fluxon number at their formation is

\begin{equation}
\Delta N_{C}=\frac{1}{2\pi}\Delta (\theta_{1}-\theta_{2}).
\end{equation}
Since $\theta_{1},\theta_{2}$ are independent as the transition
begins, we would expect, from Zurek's analysis, that
\begin{equation}
\Delta \theta_{1}=\Delta\theta_{2}\approx
 \sqrt{\frac{C}{{\bar\xi}}},
\end{equation}
whence
\begin{equation}
\Delta N_{C}\approx\frac{1}{2\pi}\sqrt{\frac{2C}{{\bar\xi}}}.
\end{equation}
Although some caution is required in the interpretation of the
experiments, which were not devised with this prediction in mind,
it seems to be supported by experiment, for which, with ${\bar\xi}
= O(10^{-1})mm$ \footnote{Corresponding to $\tau_{Q} = O(1)s$,
$\xi_{0} = O(10^{3})A$.} and $C = 0.5\times 10^{-1}mm$, say, we
would expect $\Delta N_c = O(1)$.

There is certainly no agreement, in this or any other experiment
in which defects are seen, with the thermal fluctuation density
that would be based on Eq.\ref{xiG}.

\section{The Kibble-Zurek picture for the freezing in of $\xi$ is correct.}

 Since all equations of motion have causality built into them we
should be able to confirm the first predictions Eq.\ref{xiC} and
\ref{xiZ} of Kibble and Zurek explicitly, as we shall now see.

\subsection{Condensed matter: the TDLG equation}

We assume that, for the condensed matter systems of interest to
us, the dynamics of the transition can be derived from the
explicitly {\it time-dependent} Landau-Ginzburg free energy
\begin{equation}
F(t) = \int d^{3}x\,\,\bigg(\frac{\hbar^{2}}{2m}(\nabla\phi_{a}
)^{2} +\alpha (t)\phi_{a}^{2} + \frac{1}{4}\beta
(\phi_{a}^{2})^{2}\bigg). \label{F}
\end{equation}
in which we substitute $T(t)$ for $T$ directly in (\ref{FNR}). In
(\ref{F}) $\phi = (\phi_{1} + i\phi_{2})/\sqrt{2}$ ($a=1,2$) is
the complex order-parameter field, whose magnitude determines the
superfluid density. As before, in a mean field approximation, the
chemical potential $\alpha (T)$ takes the form $\alpha (T) =
\alpha_{0}\epsilon ({\bar t})$, where $\epsilon = (T/T_c -1)$. In
a quench in which $T_c$ or $T$ changes it is convenient to shift
the origin in time, to write $\epsilon$  as
\begin{equation}
\epsilon (t) = \epsilon_{0} - \frac{t}{\tau_{Q}}\theta (t)
\label{eps}
\end{equation}
for $-\infty < t < \tau_{Q}(1 + \epsilon_{0})$, after which
$\epsilon (t) = -1$.  $\epsilon_{0} = 1 - T_{0}/T_c$ measures the
original temperature $T_{0}$ and $\tau_{Q}$ defines the quench
rate.  The quench begins at time $t = 0$ but the transition from
the normal to the superfluid phase only begins at time $t_{0} =
\epsilon_{0}\tau_{Q}$.  When it is convenient to measure time from
the onset of the transition we use the notation $\Delta t = t
-t_0$.

Motivated by Zurek's later numerical\cite{zurek2} simulations, we
adopt the time-dependent Landau-Ginzburg (TDLG) equation for $F$,
\begin{equation}
\frac{1}{\Gamma}\frac{\partial\phi_{a}}{\partial t} =
-\frac{\delta F}{\delta\phi_{a}} + \eta_{a}, \label{tdlg}
\end{equation}
where $\eta_{a}$ is Gaussian thermal noise, satisfying
\begin{equation}
\langle\eta_{a} ({\bf x},t)\eta_{b} ({\bf y},t')\rangle =
2\delta_{ab}T(t)\Gamma\delta ({\bf x}-{\bf y})\delta (t -t').
\label{noise}
\end{equation}
This is a crude approximation for $^{4}He$, and a simplified form
of a realistic description of $^{3}He$ but it is not a useful
description of QFT, as it stands.

It is relatively simple to determine the validity of Zurek's
argument since it assumes that freezing in of field fluctuations
occurs just before symmetry breaking begins. At that time the
effective potential $V(\phi ,T)$ is still roughly quadratic and we
can see later that, for the relevant time-interval $-{\bar t}\leq
\Delta t\leq {\bar t}$ the self-interaction term can be neglected
($\beta =0$).

In space, time and temperature units in which $\xi_{0} = \tau_{0}
= k_{B} =1$, Eq.\ref{tdlg} then becomes
\begin{equation}
{\dot\phi}_{a}({\bf x},t) = - [-\nabla^{2} + \epsilon (t)]\phi_{a}
({\bf x},t) +{\bar\eta}_{a} ({\bf x},t). \label{free}
\end{equation}
where ${\bar\eta}$ is the renormalised noise. The solution of the
'free'-field linear equation is straightforward, giving a Gaussian
equal-time correlation function
\begin{equation}
\langle\phi_{a} ({\bf r},t)\phi_{b} ({\bf 0},t)\rangle
=\delta_{ab}G({\bf r},t) = \int d \! \! \! / ^3 k \, e^{i {\bf k}
. {\bf r} } P(k, t). \label{diag}
\end{equation}
in which the power spectrum $P(k,t)$ has a representation in terms
of the Schwinger proper-time $\tau$ as
\begin{equation}
P(k, t) = \int_{0}^{\infty} d\tau \,{\bar T}(t-\tau/2)\,e^{-\tau
k^{2}}\,e^{-\int_{0}^{\tau} ds\,\,\epsilon (t- s/2)},
\label{lgpower}
\end{equation}
where ${\bar T}$ is the renormalised temperature. In turn, this
gives\cite{ray3}
\begin{equation}
G(r, t) = \int_{0}^{\infty} d\tau\,{\bar T}(t-\tau/2)
\,\bigg(\frac{1}{4\pi\tau}\bigg)^{3/2}
e^{-r^{2}/4\tau}\,e^{-\int_{0}^{\tau} ds\,\,\epsilon (t- s/2)}.
\label{lgcorr}
\end{equation}

For constant $\epsilon$, as happens at early times, on rescaling
in Eq.\ref{lgcorr} we recover the usual Yukawa correlator
\begin{equation}
G(r,t) = \frac{T_0}{4\pi r}\,e^{-r/\xi_{0}}, \label{yuk}
\end{equation}
where $T_0 = T_c (1 + \epsilon_0 )$ is the initial temperature.
For $\epsilon (t)$ of Eq.\ref{eps}  a saddle-point calculation
shows that, although we recover Eq.\ref{yuk} for
$r/{\bar\xi}>({\bar\xi}/\xi_{0})^{3}$ at later times, the
correlation function is then dominated by its smaller-r behaviour.
At time $t_{0} = \epsilon_{0}\tau_{0}$, when the transition
begins,  a saddle-point calculation shows that, provided the
quench is not too fast,
\begin{equation}
G(r,t_{0})\approx \frac{T_c}{4\pi r}\,e^{-a(r/{\bar\xi})^{4/3}},
\label{notyuk}
\end{equation}
where $a = O(1)$, confirming Zurek's result.

Zurek's prediction is robust, since further  calculation shows
that
 $\xi (t)$ does not vary strongly in the interval
$-{\bar t} \leq \Delta t\leq {\bar t}$, where $\Delta t =
t-t_{0}$.

\subsection{QFT: Closed time-path ensemble averaging}

For QFT the situation is rather different. In the previous
section, instead of working with the TDLG equation, we could have
worked with the equivalent Fokker-Planck equation for the
probability $p_{t}[\Phi ]$ that, at time $t>0$, the measurement of
$\phi$ will give the function $\Phi ({\bf x})$. When solving the
dynamical equations for a hot quantum field  it is convenient to
work with probabilities from the start.

Take $t=0$ as our starting time for the evolution of the complex
field $\phi = (\phi_1 +i\phi_2)/\sqrt{2}$.  Suppose that, at this
time, the system is in a pure state, in which the measurement of
$\phi$ would give $\Phi_0({\bf x})$. That is:-
\begin{equation}
\hat{\phi}(t=0,{\bf x}) | \Phi_0,t=0 \rangle = \Phi_0 | \Phi_0,t=0
\rangle.
\end{equation}
The probability $p_{t}[\Phi]$ that, at time $t_f>0$, the
measurement of $\phi$ will give the value $\Phi$ is $p_{t}[\Phi] =
|\Psi_{0}|^2$, where $\Psi_{0}$ is the state-functional with the
specified initial condition.  As a path integral
\begin{equation}
\Psi_{0}  = \int_{\phi(0) = \Phi_0}^{\phi(t) = \Phi} {\cal D} \phi
\, \exp \biggl \{ i S_t [\phi] \biggr \},
\end{equation}
where $S_t [\phi]$ is the (time-dependent) action that describes
how the field $\phi$ is driven by the environment
 and spatial and field
labels have been suppressed (e.g. ${\cal D}\phi ={\cal
D}\phi_1{\cal D}\phi_2 )$. Specifically, for $t > 0$ the action
for the field is taken to be
\begin{equation}
S_t [\phi] = \int dx \biggl ( \frac{1}{2} \partial_{\mu} \phi_{a}
\partial^{\mu} \phi_{a} - \frac{1}{2} m^{2}(t) \phi_{a}^2 -
\frac{1}{4} \lambda (\phi_{a}^2)^2 \biggr ). \label{St}
\end{equation}
where $m(t)$ describes the evolution of the action under external
influences, to which the field responds.

It follows that $p_{t}[\Phi]$ can be written in the closed
time-path form
\begin{equation}
p_{t}[\Phi] = \int_{\phi_{\pm}(0) = \Phi_0}^{\phi_{\pm}(t) = \Phi}
{\cal D} \phi_+  {\cal D} \phi_- \, \exp \biggl \{ i \biggl ( S_t
[\phi_+]-S_t [\phi_-] \biggr ) \biggr \},
\end{equation}
where ${\cal D}\phi_{\pm} ={\cal D}\phi_{\pm ,1}{\cal D}\phi_{\pm
,2} $. Instead of separately integrating $\phi_{\pm}$ along the
time paths $0 \leq t \leq t_f$, the integral can be interpreted as
time-ordering of a field $\phi$ along the closed path $C_+ \oplus
C_-$ where $\phi =\phi_+$ on $C_+$ and $\phi= \phi_-$ on $C_-$.
 When we extend
the contour from $t_f$ to $t= \infty$ either $\phi_+$ or $\phi_-$
is an equally good candidate for the physical field, but we choose
$\phi_+$.
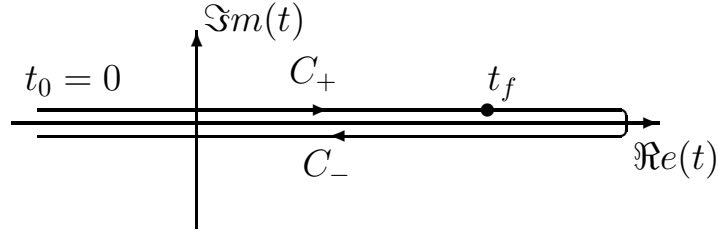
\begin{figure}[htb]
\begin{center}
\setlength{\unitlength}{0.5pt}
\begin{picture}(495,200)(35,600)
\put(260,640){\makebox(0,0)[lb]{\large $C_-$}}
\put(185,750){\makebox(0,0)[lb]{\large $\Im m(t)$}}
\put(250,710){\makebox(0,0)[lb]{\large $C_+$}}
\put(510,645){\makebox(0,0)[lb]{\large $\Re e (t)$}}
\put(400,705){\makebox(0,0)[lb]{\large $t_f$}} \put(
50,705){\makebox(0,0)[lb]{\large $t_0=0$}} \thicklines
\put(400,690){\circle*{10}} \put( 40,680){\vector( 1, 0){490}}
\put( 60,690){\vector( 1, 0){220}} \put(280,690){\line( 1,
0){220}} \put(500,680){\oval(10,20)[r]} \put(500,670){\vector(-1,
0){220}} \put(280,670){\line(-1, 0){220}}
\put(180,600){\vector( 0, 1){150}}
\end{picture}
\end{center}
\caption{The closed timepath contour $C_+ \oplus C_-$.}
\end{figure}

The choice of a pure state at time $t=0$ is too simple to be of
any use. As we said earlier, we assume that $\Phi$ is Boltzmann
distributed at time $t=0$ at an effective temperature of $T_0 =
\beta_0^{-1}$ according to the  Hamiltonian $H[\Phi]$
corresponding to the action $S[\phi ]$, in which $\phi$ is taken
to be periodic in imaginary time with period $\beta_0$. We now
have the explicit form for $p_{t}[\Phi]$,
\begin{equation}
p_{t} [ \Phi] = \int_B {\cal D} \phi \, e^{i S_C [\phi]} \, \delta
[ \phi_+ (t_f) - \Phi ],
\end{equation}
 written as the time ordering of a
single field along the contour $C=C_+ \oplus C_- \oplus C_3$,
extended to include a third imaginary leg, where $\phi$ takes the
values $\phi_+$, $\phi_-$ and $\phi_3$ on $C_+$, $C_-$ and $C_3$
respectively, for which $S_C$ is $S[\phi_+]$, $S[\phi_-]$ and
$S_0[\phi_3]$.
\begin{figure}[htb]
\begin{center}
\setlength{\unitlength}{0.5pt}
\begin{picture}(495,280)(35,480)
\put( 70,565){\makebox(0,0)[lb]{\large $C_3$}}
\put(260,640){\makebox(0,0)[lb]{\large $C_-$}}
\put(185,750){\makebox(0,0)[lb]{\large $\Im m(\tau)$}}
\put(250,710){\makebox(0,0)[lb]{\large $C_+$}}
\put(510,645){\makebox(0,0)[lb]{\large $\Re e (\tau)$}}
\put(400,705){\makebox(0,0)[lb]{\large $t_f$}} \put(
50,705){\makebox(0,0)[lb]{\large $t_0$}}
\put(70,490){\makebox(0,0)[lb]{\large $t_0-i \beta_0$}}
\thicklines \put(400,690){\circle*{10}} \put( 40,680){\vector( 1,
0){490}} \put( 60,690){\vector( 1, 0){220}} \put(280,690){\line(
1, 0){220}} \put(500,680){\oval(10,20)[r]}
\put(500,670){\vector(-1, 0){220}} \put(280,670){\line(-1,
0){220}} \put( 60,670){\vector( 0,-1){110}} \put( 60,560){\line(
0,-1){ 60}} \put(180,480){\vector( 0, 1){280}}
\end{picture}
\end{center}
\caption{A third imaginary leg}
\end{figure}
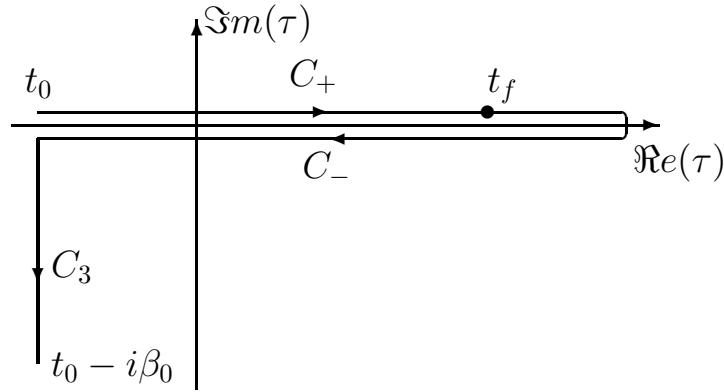

To demonstrate how we can average without having to calculate
$p_t[\Phi]$ explicitly we see that $G_{ab}(|{\bf x} -{\bf x} '|;t)
= \langle\Phi_a ({\bf x})\Phi_b ({\bf x}')\rangle_{t}$ is given by
\begin{equation}
G_{ab}(|{\bf x} -{\bf x} '|;t) = \langle\phi_a ({\bf x},t)\phi_b
({\bf x}',t)\rangle, \label{wight}
\end{equation}
the equal-time thermal Wightman function with the given thermal
boundary conditions. Because of the non-equilibrium time evolution
there is no time translation invariance in the double time label.

\subsection{QFT: the free roll}

Fortunately, as for the condensed matter case, the interval
$-{\bar t} \leq \Delta t\leq {\bar t}$ occurs in the {\it linear}
regime, when the self-interactions are unimportant. The relevant
equation for constructing the correlation functions of this
one-field system is now the second-order equation
\begin{equation}
\frac{\partial^{2}\phi_a}{\partial t^{2}} = -\frac{\delta
F}{\delta\phi_a}, \label{op}
\end{equation}
for $F$ of Eq.\ref{FR}. This is solvable in terms of the mode
functions $\chi^{\pm}_{k}(t)$, identical for $a=1,2$, satisfying
\begin{equation}
\Biggl [ \frac{d^2}{dt^2} + {\bf k}^2 + m^2(t) \Biggr
]\chi^{\pm}_{k}(t)  =0, \label{mode}
\end{equation}
subject to $\chi^{\pm}_{k}(t)
=
e^{\pm i\omega_{in}t}$ at $t\leq 0$, for incident frequency
$\omega_{in} = \sqrt{{\bf k}^{2} +\epsilon_{0} M^{2}}$, for
$m^2(t) = \epsilon (t)M^{2}$, where $\epsilon (t)$ is
parameterised as for the  TDLG equation above.  This corresponds
to a temperature quench from an initial state of thermal
equilibrium at temperature $T_{0}>T_c $, where $(T_{0}/T_c -1) =
\epsilon_{0}$. There is no reason to take $\epsilon_{0}$ small.
The diagonal correlation function $G(r, t)$ of Eq.\ref{diag} is
given as the equal-time propagator
\begin{eqnarray}
G(r, t)&=&\int d \! \! \! / ^3 k \, e^{i {\bf k} . {\bf x} }
\chi^{+}_{k}(t) \chi^{-}_{k}(t)C(k)
\\
\nonumber &=& \frac{1}{2\pi^{2}}\int dk\,k^{2}\frac{\sin kr}{kr}
\chi^{+}_{k}(t) \chi^{-}_{k}(t)C(k), \label{modesum}
\end{eqnarray}
where $ C(k) =\coth(\omega_{in} (k)/2T_0 )/2\omega_{in}(k)$
encodes the initial conditions.

An exact solution can be given\cite{bowick} in terms of Airy
functions. Dimensional analysis shows that, on ignoring the
k-dependence of $C(k)$, appropriate for large $r$ (or small $k$),
$\xi_{eq}({\bar t})$ of Eq.\ref{xiC} again sets the scale of the
equal-time correlation function. Specifically,
\begin{equation}
G(r,\Delta t =0)\propto\int d\kappa\,\frac{\sin\kappa
(r/{\bar\xi})}{\kappa (r/{\bar\xi})}\,F(\kappa ), \label{GQFT}
\end{equation}
where $F(0) = 1$ and $F(\kappa )\sim \kappa^{-3}$ for large
$\kappa$.
 Kibble's insight is
correct, at least for this case of a single (uncoupled) field.

\section{However, Defect Densities do not Determine ${\bar\xi}$ Directly (and Vice-Versa)}

We have seen that there is no reason to disbelieve the causal
arguments of Kibble for QFT and Zurek for condensed matter as to
the correlation length ${\bar\xi}$ at the onset of the transition.
The excellent agreement with the $^{3}He$ experiments suggests
that, for condensed matter, this length does, indeed, characterise
vortex separation at the time when the defects form.

However, if we take the Lancaster experiment at face value, this
cannot always be the case.  This is not surprising.  Numerical
simulations of TDLG equations, by Zurek himself, and the related
Langevin equations for non-equilibrium statistical fields show
that, at early times after a transition, the field fluctuations
remain on very small scales.
 In reality, different frequency
modes freeze at different times, and the causal argument only
applies strictly to the long wavelength modes. Of course, field
ordering is controlled by the long wavelength modes, but we would
like a more complete description of the freezing in of the field,
if we are to make the second step of relating this to defect
structure in the field . That is, although the field is correlated
on long scales ${\bar\xi}$, this does not preclude structure on
very much shorter scales.  Moreover, it is this structure, that we
might wish to think of as proto-defects, that will evolve into the
defects, in this case vortices, once the transition is complete.
Inspection of the field at early times shows no obvious relation
between the separation of these proto-defects and ${\bar\xi}$. The
question then becomes one of why the $^{3}He$ experiments should
be in agreement, rather than why the $^{4}He$ experiments are not.

\subsection{Classical vortices in condensed matter and QFT}

The $O(2)$ string is, classically, a tube of false vacuum of
radius $O(M^{-1})$ whose core is characterised by a line of string
zeroes. Its winding number $n\in Z$ is the change in phase of the
complex field around the core (in units of $2\pi$).  We shall only
consider strings with $|n| = 1$ since all others can be considered
as multiple zeroes.

It would be foolish to estimate the probability of finding
vortices directly from $p_{t}[\Phi ]$. A starting-point for
counting vortices in superfluids is to count line zeroes.
 Not all line zeroes are candidates for vortices since zeroes occur on all
 scales. However, a starting-point for
counting vortices in superfluids is to count line zeroes
 of an
appropriately coarse-grained field, in which structure on a scale
smaller than $\xi_{0}$, the classical vortex size, is not
present\cite{popov}. This is also the unstated basis of the
numerous numerical simulations\cite{tanmay} of cosmic string
networks built from Gaussian fluctuations (but see
\cite{gleiser}).

Even then, there are several prerequisites before line zeroes can
be identified with vortex cores, and $n_{zero}(t)$ with
$n_{def}(t)$.
\begin{itemize}
\item
The field, on average must have achieved its symmetry-broken
ground-state equilibrium value
\begin{equation}
\langle |\phi |^{2}\rangle = \alpha_{0}/\beta\,\,\mbox{or}
M^{2}/\lambda , \label{eq}
\end{equation}
 non-perturbatively large (in $\beta$)
This, in itself, is sufficient to show that the causal time ${\bar
t}$ is {\it not} the time to begin looking for defects, since
$\langle |\phi |^{2}\rangle$ is small at this time.
\item
Only when  $\partial n_{zero}/\partial l$ is small in comparison
to $n_{zero}/l$ at $l = \xi_{0}$ will the line-zeroes have the
small-scale non-fractal nature of classical defects, although
defects
 may behave like random walks on larger scales.  As the power in the long
wavelength modes increases the 'Bragg' peak develops in
$k^{2}G(k,t)$, moving in towards $k=0$.  This condition then
becomes the condition that the peak dominates its tail.
\item
The energy in field gradients should be commensurate with the
energy in classical vortices with the same density as that of line
zeroes.
\end{itemize}

We stress that these are necessary, but not sufficient, conditions
for classical vortices.  In particular,  only the full
nonlinearity of the system can establish classical profiles.  We
will term such zeroes as satisfy these conditions, proto-vortices.
In fact, most (but not all\cite{calzetta}) numerical lattice
simulations cannot distinguish between proto-vortices and
classical vortices.

Whereas the above are equally true for condensed matter and QFT
there are further complications peculiar to QFT. In particular, in
QFT we need to consider the whole density matrix $\langle\Phi '
|\rho (t)|\Phi \rangle$ rather than just the diagonal elements
$p_{t}[\Phi ] = \langle\Phi |\rho (t)|\Phi \rangle$.  Classicality
is understood in terms of 'decoherence', manifest most simply by
the approximate diagonalisation of the reduced density matrix on
coarse-graining. By this we mean the separation of the whole into
the 'system', and its 'environment' whose degrees of freedom are
integrated over, to give a reduced density matrix.  The
environment can be either other fields with which our scalar is
interacting or even the short wavelength modes of the scalar field
itself \cite{muller2,lombardo}. When interactions are taken into
account this leads to quantum noise and dissipation.

In the Gaussian approximations for QFT that we shall adopt here,
with $\langle\Phi \rangle = 0$, integrating out short wavelengths
with $k>l^{-1}$ is just equivalent to a momentum cut-off at the
same value. This gives neither noise nor dissipation and
diagonalisation does not occur. Nonetheless, from our viewpoint of
counting line-zeroes, fluctuations are still present when $l =
O(M^{-1})$ that prevent us from identifying line-zeroes with
proto-vortices easily.

For all these caveats, there are other symptoms of classical
behaviour in QFT once $G_{l}(0;t)$ is non-perturbatively large.
Instead of a field basis, we can work in a particle basis and
measure the particle production as the transition proceeds. We
shall see that, in the Gaussian approximation, $n_{zero}$, the
density of line zeroes, is given in terms of the moments of
$G_{l}(r,t)$. Whether we expand with respect to the original Fock
vacuum or with respect to the adiabatic vacuum state, the presence
of a non-perturbatively large peak in $k^2 G(k;t)$ at $k = k_0$
signals a non-perturbatively large occupation number
$N_{k_{0}}\propto 1/\lambda$ of particles at the same wavenumber
$k_0$\cite{boyanovsky}. With $n_{zero}$ of order $k_{0}^{2}$ this
shows that the long wavelength modes can now begin to be treated
classically.

From a slightly different viewpoint, the Wigner functional only
peaks about the classical phase-space trajectory once the power is
non-perturbatively large\cite{guth,muller}. More crudely, the
diagonal density matrix elements are only then significantly
non-zero for non-perturbatively large field configurations
$\phi\propto\lambda^{-1/2}$ like vortices.

\subsection{Line-zero density}

Suppose, at some time, that the field has line zeroes ${\bf
R}_{n}(s)$, where $n = 1,2,..$ labels the zero, and $s$ measures
the length along it.  As a result the {\it topological line
density} of zeroes ${\vec{\rho}} (\bf r)$ can be
defined\cite{halperin,maz} by
\begin{equation}
\vec{\rho}({\bf x}) = \sum_{n}\int ds \frac{d{\bf R}_{n}}{ds}
\delta^{3} [{\bf x} - {\bf R}_{n}(s)]. \label{corrr}
\end{equation}
In (\ref{corrr}) $ds$ is the incremental length along the line of
zeroes ${\bf R}_{n}(s)$ ($n$=1,2,.. .) and $\frac{d{\bf
R}_{n}}{ds}$ is a unit vector pointing in the direction which
corresponds to positive winding number.

It follows that, in terms of the zeroes of the field $\Phi ({\bf
x})$, $\rho_{i}({\bf x})$ can be written as
\begin{equation}
\rho_{i}({\bf x}) = \delta^{2}[\Phi ({\bf
x})]\epsilon_{ijk}\partial_{j} \Phi_{1}({\bf x})
\partial_{k}\Phi_{2}({\bf x}), \label{rho3}
\end{equation}
where $\delta^{2}[\Phi ({\bf x})] = \delta[\Phi_{1} ({\bf x})]
\delta[\Phi_{2} ({\bf x})]$. The coefficient of the
$\delta$-function in (\ref{rho3}) is the Jacobian of the more
complicated transformation from line zeroes to field zeroes. What
we want is not this, but the {\it total line density}
$\bar{\rho}({\bf x})$,
\begin{equation}
\bar{\rho_{i}}({\bf x}) = \delta^{2}[\Phi ({\bf
x})]|\epsilon_{ijk}\partial_{j} \Phi_{1}({\bf x})
\partial_{k}\Phi_{2}({\bf x})|. \label{rhobar}
\end{equation}

The vanishing field expectation value and the  independence of the
field and its derivatives
\begin{equation}
\langle\Phi_{a}({\bf x})\rangle = 0 = \langle\Phi_{a}({\bf
x})\partial_{j}\Phi_{b}({\bf x})\rangle , \label{gg4}
\end{equation}
imply $\langle\rho_{j}({\bf x})\rangle = 0$ {\it i.e.} an equal
likelihood of a string line-zero or an antistring line-zero
passing through an infinitesimal area. However,
\begin{equation}
n(t) = \; \langle\bar{\rho_{i}}({\bf x})\rangle_{t} \; > 0
\label{n3}
\end{equation}
and measures the {\it total} line-zero density in the direction
$i$, without regard to string orientation.  The isotropy of the
initial state guarantees that $n$ is independent of the direction
$i$.

Whereas the  correlation length ${\bar\xi} =\xi_{eq}({\bar t})$
depends on the {\it long-distance} behaviour of $G(r,t)$, this is
not the case for the line-zero density $n_{zero}$. In our Gaussian
approximation\cite{halperin,maz} of the previous section it is
determined completely by the {\it short-distance} behaviour of
$G(r, t)$ as
\begin{eqnarray}
n_{zero}(t)& = &\frac{-1}{2\pi}\frac{G''(0, t)}{G(0, t )}
\label{ndeff}
\\
\nonumber &=& -\frac{1}{2\pi}f''(0, t), \,\,\,\,\mbox{where}\,\,
f(r, t) = \frac{G(r, t)}{G(0, t)}
\end{eqnarray}

Some caution is necessary. Since thermal and quantum fluctuations
will give rise to zeroes on all scales, neither $G(0,t)$ nor its
derivatives exist because of ultraviolet divergences. For the
moment, we put in a cutoff $l = O(\xi_{0})$ by hand, as
\begin{equation}
G_{l}(r, t) = \int d \! \! \! / ^3 k\, e^{i{\bf k}.{\bf x}}G(k,
t)\,e^{-k^{2}l^{2}}. \label{Gl}
\end{equation}
We note that the inclusion of a cut-off does not affect the
long-distance correlation ${\bar\xi}$, which depends essentially
on the weighting of the nearest singularities of $G(k, t)$ in the
complex $k$-plane.

In the linear regime everything is calculable. If we define the
line-zero separation $\xi_{zero}(l)$ by
\begin{equation}
n_{zero}(l, t) = \frac{1}{2\pi\xi_{zero}(l, t)^{2}} \label{nzdef}
\end{equation}
it is apparent that $\xi_{zero}(t)$ has little, if anything, to do
with ${\bar\xi}$ directly.

For non-Gaussian fields the situation is much more complicated.
However, as long as there is a dominant wavenumber $k_0$ in
$G(k;t)$ this sets a length scale $\xi\approx k^{-1}_0$ that
characterises vortex separation.  At the level of density
calculations, as distinct from length distributions, this is
essentially all that is needed.

We can already anticipate a mismatch between the defect density
and the long-range correlation length that could derail the
Kibble-Zurek predictions.

\subsection{Annular experiments require a new correlation length}

We conclude with a brief discussion of annular experiments, that
have no counterpart in QFT. To the extent that an annulus can be
treated as a one-dimensional system there has been good numerical
work. However, to date there are no reliable calculations for
quenching in an annulus that take realistic boundary conditions
into account. Nonetheless, we have some hints as how to proceed.
Consider a circular path in the bulk fluid (in the 1-2 plane),
circumference $C$, the boundary of a surface $S$. For given field
configurations $\phi_{a}({\bf x})$ the phase change $\theta_{C}$
along the path can be expressed as the surface integral
\begin{equation}
\theta_{C} = 2\pi\int_{{\bf x}\in S} d^{2}x\,\,\rho_{3}({\bf x}),
\end{equation}

Again we quench from an initial state with no rotation. The
variance in the phase change around $C$, $\Delta\theta_{C}$ is
determined from
\begin{equation}
(\Delta\theta_{C})^{2} =4\pi^{2}\int_{{\bf x}\in
S}d^{2}x\int_{{\bf y}\in S}d^{2}y\,\langle\rho_3 ({\bf x})\rho_3
({\bf y})\rangle_{t}.
\end{equation}
On using the conservation of charge it is not difficult to show,
from the results of\cite{halperin,maz} that, for Gaussian fields,
with momentum cut-off at $k = l^{-1}$, $\Delta\theta_{C}$
satisfies
\begin{equation}
(\Delta\theta_{C})^{2} =-\int_{{\bf x}\not\in S}d^{2}x\int_{{\bf
y}\in S}d^{2}y\,\,{\cal C}_{l}(|{\bf x}-{\bf y}|,t), \label{per}
\end{equation}
where ${\bf x}$ and ${\bf y}$ are in the plane of $S$, and
\begin{equation}
{\cal C}_{l}(r,t) = \frac{1}{r}\frac{\partial}{\partial
r}\bigg(\frac{f\prime_{l}^{2}(r,t)}{1 - f_{l}^{2}(r,t)}\bigg).
\label{C2}
\end{equation}

Since $G_{l}(r,t)$ is short-ranged
 ${\cal C}_{l}(r,t)$ is short-ranged also.
With ${\bf x}$ outside $S$, and ${\bf y}$ inside $S$, all the
contribution to $(\Delta\theta_{C})^{2}$ comes from the vicinity
of the boundary of $S$, rather than the whole area. That is, if we
removed all fluid except for a strip from the neighbourhood of the
contour $C$ we would still have the same result.  This supports
the assertion by Zurek that the correlation length for phase
variation in bulk fluid is also appropriate for annular flow.  The
purpose of the annulus (more exactly, a circular capillary of
circumference $C$ with radius $l\ll C$) is to stop this flow
dissipating into the bulk fluid.

More precisely, suppose that $C\gg\xi (t)$. Then, if we take the
width $2l$ of the strip around the contour to be larger than the
correlation length of ${\cal C}_{l}(r,t)$, Eq.\ref{per} can be
written as
\begin{equation}
(\Delta\theta_{C})^{2}\approx -2
\,C\int_{0}^{\infty}dr\,r^{2}\,\,{\cal C}_{l}(r,t). \label{per2}
\end{equation}
The linear dependence on $C$ is purely a result of Gaussian
fluctuations.

Insofar as we can identify the bulk correlation with the annular
correlation, instead of Eq.\ref{v1}, we have
\begin{equation}
\Delta v = \frac{\hbar}{m}\sqrt{\frac{1}{C\xi_{s}(t)}}. \label{v3}
\end{equation}
The step length $\xi_{s}(t)$ is given by
\begin{equation}
\frac{1}{\xi_{s}(t)} =
2\int_{0}^{\infty}dr\frac{f\prime_{l}^{2}(r,t)}{1 -
f_{l}^{2}(r,t)}.
 \label{step}
\end{equation}

 If we quench in an annular capillary of radius $l$ much
smaller than its circumference, we are, essentially,
coarsegraining to that scale.  That is, the observed variance in
the flux along the annulus is $\pi l^{2}\Delta v$ for $\Delta v$
averaged on a scale $l$. We make the approximation that that is
the {\it major} effect of quenching in an annulus. This cannot be
wholly true, but it is plausible if the annulus is not too narrow
for boundary effects to be important.

\section{Nonetheless, Zurek's Predictions are Correct When
Thermal Fluctuations are Small}

An integral part of the Zurek/Kibble predictions for $n_{def}$ is
their universality.  The ratio Eq.\ref{ndeff} for $n_{zero}$ gives
intimations as to how this could happen, insofar as the the
specific effects of the interactions can be subsumed into
prefactors that cancel. Although, in the context of Gaussian
fluctuations, it can only be approximate, it is part of a general
truth that defect density only depends on very limited attributes
of the fluctuation power spectra.

We can see this in the models that we have introduced. The TDGL
equation is highly dissipative, whereas the QFT equation is not,
as it stands. Nonetheless, in each case the field will become
ordered after the onset of the transition by the growth of
unstable long wavelength modes. How this happens depends on the
specifics of the equations. We would therefore expect the field
behaviour as the transition is being implemented to differ in
detail for condensed matter and QFT, but only weakly affecting the
initial line-zero density, and it s ability to describe vortices.

\subsection{TDLG condensed matter: vortex densities}

We begin with condensed matter, which we will find to be easier.
As the system evolves away from the transition time, the free
equation Eq.\ref{free} ceases to be valid, to be replaced by the
full equation
\begin{equation}
{\dot\phi}_{a}({\bf x},t) = - [-\nabla^{2} + \epsilon
(t)+{\bar\beta}|\phi ({\bf x},t)|^{2}]\phi_{a} ({\bf x},t)
+{\bar\eta}_{a} ({\bf x},t), \label{full}
\end{equation}
where ${\bar\beta}$ is the rescaled coupling.

In order to retain some analytic understanding of the way that the
density is such an ideal quantity to make predictions for, we
adopt the approximation of preserving Gaussian fluctuations by
linearising the self-interaction as
\begin{equation}
{\dot\phi}_{a}({\bf x},t) = - [-\nabla^{2} + \epsilon_{eff}
(t)]\phi_{a} ({\bf x},t) +{\bar\eta}_{a} ({\bf x},t),
\end{equation}
where $\epsilon_{eff} $ contains a (self-consistent) term
$O({\bar\beta}\langle |\phi |^{2}\rangle)$. Additive
renormalisation is necessary, so that $\epsilon_{eff}\approx
\epsilon$, as given earlier, for $t\leq t_{0}$.

Self-consistent linearisation is the standard approximation in
non-equilibrium QFT\cite{boyanovsky}, but is not strictly
necessary here, since numerical simulations that identify line
zeroes of the field can be made that use the full
self-interaction\cite{zurek2}. However, to date none address the
questions we are posing here exactly, and until then there is
virtue in analytic approximations provided they are not taken too
seriously.

 The solution for $G(r,t)$ is a straightforward generalisation of
\ref{lgcorr},
\begin{equation}
G(r,t) =\int_{0}^{\infty} d\tau \,{\bar T}(t-\tau /2)
\bigg(\frac{1}{4\pi\tau}\bigg)^{3/2}
e^{-r^{2}/4\tau}\,e^{-\int_{0}^{\tau} ds\,\,\epsilon_{eff} (t-
s/2)}.
\end{equation}
Putting in the momentum cutoff $k^{-1}> l =\bar{l}\xi_{0}
=O(\xi_{0} )$ of Eq.\ref{Gl} by hand corresponds to damping the
singularity in $G(r,t)$ at $\tau = 0$ as\cite{ray3}
\begin{equation}
G_{l}(r,t)=\int_{0}^{\infty}\frac{ d\tau\,{\bar T}(t-\tau
/2)}{[4\pi (\tau + {\bar l}^{2})]^{3/2} }
e^{-r^{2}/4\tau}\,e^{-\int_{0}^{\tau} ds\,\,\epsilon_{eff} (t-
s/2)},
\end{equation}
making $G_{l}(0,t)$ finite. We stress that, for $t\approx t_{0}$,
the correlation length ${\xi}$ remains $O({\bar\xi})$, {\it
independent} of $l$.

Assuming a {\it single} zero of $\epsilon_{eff} (t)$ at $t =
t_{0}$, at $r=0$ the exponential in the integrand peaks at $\tau
={\bar\tau} = 2(t-t_{0})$.  Expanding about ${\bar\tau}$ to
quadratic order gives
\begin{equation}
G_{l}(0,t)\approx {\bar
T}_{c}\,e^{2\int_{t_{0}}^{t}du\,|\epsilon_{eff} (u)|
}\int_{0}^{\infty} \frac{ d\tau\,e^{-(\tau -
2(t-t_{0}))^{2}|\epsilon '(t_{0})|/4}}{[4\pi (\tau + {\bar
l}^{2})]^{3/2} } .
\end{equation}
For times $t > \epsilon_{0}\tau_{Q}$ we see that, as the
unfreezing occurs, long wavelength modes with $k^{2} < t/\tau_Q -
\epsilon_{0}$ grow exponentially.

The effect of the back-reaction is to stop the growth of
$G_{l}(0,t)-G_{l}(0,t_{0})= \langle |\phi |^{2}\rangle_{t}-\langle
|\phi |^{2}\rangle_{0}$ at its symmetry-broken value
${\bar\beta}^{-1}$ in our dimensionless units. A necessary
condition for this is $\lim_{u\rightarrow\infty}\epsilon_{eff} (u)
= 0$. That is, we must choose
\begin{equation}
\epsilon_{eff}(t) = \epsilon (t)  +
{\bar\beta}(G_{l}(0,t)-G_{l}(0,t_{0})),
\end{equation}
thereby preserving Goldstone's theorem.

Beyond that, what is remarkable in this approximation is that the
density of line zeroes uses {\it no} property of the self-mass
contribution to $\epsilon_{eff}(t)$, self-consistent or otherwise.
With
\begin{equation}
-G''(0, t) = \frac{1}{2}\int_{0}^{\infty} \frac{d\tau}{\tau}
\,{\bar T}(t-\tau
/2)\bigg(\frac{1}{4\pi\tau}\bigg)^{3/2}\,e^{-\int_{0}^{\tau}
ds\,\,\epsilon_{eff} (t- s/2)}. \label{lgcorr3}
\end{equation}
all prefactors in $n_{zero}$ cancel\footnote{Our ignoring
prefactors in \cite{ray} was fortuitous, leaving our conclusions
obtained there unaffected.}, to give\cite{ray3,ray}
\begin{equation}
n_{zero}(t) = \frac{1}{4\pi}\frac{\int_{0}^{\infty}
\frac{d\tau}{(\tau + {\bar l}^{2})^{5/2}} \,e^{-(\tau -
2(t-t_{0}))^{2}/4{\bar t}^{2}}} {\int_{0}^{\infty}
\frac{d\tau}{(\tau + {\bar l}^{2})^{3/2}} \,e^{-(\tau - 2(t-
t_{0}))^{2}/4{\bar t}^{2}}}
\end{equation}
on using the definition $\tau_{Q} = {\bar t}^{2}$ in natural
units.

At $t=t_{0}$ both numerator and denominator are dominated by the
short wavelength fluctuations at small $\tau$. Even though the
field is correlated over a distance ${\bar\xi}\gg l$ the density
of line zeroes $n_{zero} = O(l^{-2})$ depends entirely on the
scale at which we look. In no way would we wish to identify these
line zeroes with prototype vortices. However, as time passes the
peak of the exponential grows and $n_{zero}$ becomes increasingly
insensitive to $l$. How much time we have depends on the magnitude
of ${\bar\beta}$, since once $G(0,t)$ has reached this value it
stops growing. Since $G(0,t) = O(\exp(((t-t_{0})/{\bar t})^{2})$
at early times the backreaction is implemented extremely rapidly.
We can estimate the time $t^{*}$ at which this happens by
substituting $\epsilon (u)$ for $\epsilon_{eff} (u)$ in the
expression for $G_{l}(0,t)$ above.

For $t>t^{*}$ the equation for $n_{zero}(t)$ is not so simple
since the estimate above, based on a single isolated zero of
$\epsilon_{eff} (t)$, breaks down because of the approximate
vanishing of $\epsilon_{eff} (t)$ for $t>t^{*}$.  A more careful
analysis shows that $G_{l}(0,t)$ can be written as
\begin{equation}
G_{l}(0,t)\approx \int_{0}^{\infty} \frac{ d\tau\,{\bar T}(t-\tau
/2)}{[4\pi (\tau + {\bar l}^{2})]^{3/2} } {\bar G}(\tau,t),
\end{equation}
where ${\bar G}(\tau,t)$ has the same peak as before at $\tau =
2(t-t_{0})$, in the vicinity of which
\begin{equation}
{\bar G}(\tau,t) = e^{2\int_{t_{0}}^{t}du\,|\epsilon_{eff}
(u)|}\,e^{-(\tau - 2(t-t_{0}))^{2}/4{\bar t}^{2}},
\end{equation}
but ${\bar G}(\tau,t)\cong 1$ for $\tau < 2(t-t^{*})$. Thus, for
$\tau_{Q}\gg\tau_{0}$, $G_{l}(0,t)$ can be approximately separated
as
\begin{equation}
G_{l}(0,t)\cong G_{l}^{UV}(t) + G^{IR}(t),
\end{equation}
where
\begin{equation}
G_{l}^{UV}(t)= {\bar T}(t)\,\int_{0}^{\infty} d\tau\,/[4\pi (\tau
+ {\bar l}^{2})]^{3/2}
\end{equation}
 describes  the
scale-{\it dependent} short wavelength thermal noise, proportional
to temperature, and
\begin{equation}
G^{IR}(t) =\frac{{\bar T}_{c}}{(8\pi(t-t_{0}))^{3/2}}
\,\int_{-\infty}^{\infty}d\tau {\bar G}(\tau,t)
\end{equation}
describes the scale-{\it independent}, temperature independent,
long wavelength fluctuations. A similar decomposition
$G\prime\prime_{l}(0,t)\cong G\prime\prime_{l}^{UV}(t) +
G\prime\prime^{IR}(t)$ can be performed as
\begin{equation}
G\prime\prime_{l}^{UV}(t)= 2\pi{\bar T}(t)\,\int_{0}^{\infty}
d\tau\,/[4\pi (\tau + {\bar l}^{2})]^{5/2}
\end{equation}
and
\begin{equation}
G\prime\prime^{IR}(t) =\frac{4\pi{\bar
T}_{c}}{(8\pi(t-t_{0}))^{5/2}} \,\int_{-\infty}^{\infty}d\tau
{\bar G}(\tau,t).
\end{equation}

In particular, $G\prime\prime^{IR}(t)/G^{IR}(t)= O(t^{-1})$.

Firstly, suppose that, for $t\geq t^*$,
  $ G^{IR}(t)\gg G_{l}^{UV}(t)$ and  $
G\prime\prime^{IR}(t)\gg G\prime\prime_{l}^{UV}(t)$, as would be
the case for a temperature quench ${\bar T}(t)\rightarrow 0$.
 Then, with little thermal noise, we have widely separated
line zeroes, with density $n_{zero}(t)\approx
-G\prime\prime^{IR}(t)/2\pi G^{IR}(t)$. With $\partial
n_{zero}/\partial l$ small in comparison to $n_{zero}/l$ at $l =
\xi_{0}$ we identify such essentially non-fractal line-zeroes with
prototype vortices, and $n_{zero}$ with $n_{def}$.  Of course, we
require non-Gaussianity to create true classical energy profiles.
Nonetheless, the Halperin-Mazenko result may be well approximated
for a while even when the fluctuations are no longer
Gaussian\cite{calzetta}. This is supported by the observation
that, once the line zeroes have straightened on small scales at
$t>t^*$, the Gaussian field energy, largely in field gradients, is
\begin{equation}
{\bar F}\approx\langle\int_{V}
d^{3}x\,\frac{1}{2}(\nabla\phi_{a})^{2}\rangle= -VG''(0,t),
\end{equation}
where $V$ is the spatial volume. This matches the energy
\begin{equation}
{\bar E}\approx V n_{def}(t)(2\pi G(0,t)) = -VG''(0,t)
\end{equation}
possessed by a network of classical global strings with density
$n_{zero}$, in the same approximation of cutting off their
logarithmic tails.

 For times $t>t^{*}$
\begin{equation}
n_{zero}(t)\approx \frac{{\bar t}}{8\pi (t-t_{0})
}\frac{1}{\xi_{0}^{2}}\sqrt{\frac{\tau_{0}}{\tau_{Q}}},
\end{equation}
the solution to Vinen's equation\cite{vinen}
\begin{equation}
\frac{\partial n_{zero}}{\partial t} = -\chi_{2}\frac{\hbar}{m}
n_{zero}^{2},
\end{equation}
 where
$\chi_{2} = 4\pi\hbar\Gamma = 4\pi\hbar /\alpha_{0}\tau_{0}= 4\pi$
if, as earlier, we motivated $\tau_{0}$ from the Gross-Pitaevskii
equation, in which $\alpha_{0}\tau_{0} =\hbar$. More
realistically, we find $\chi_{2} > 4\pi$ both for $^{4}He$ and
$^{3}He$. Taking $\tau_{0}\approx 8.0\times 10^{-12}s$ and
$\xi_{0}\approx 5.6\AA$) in the mean-field approximation for
$^{4}He$ gives $\chi_{2}\approx 5\times 4\pi$. For $^{3}He$ (with
$\xi_{0} \approx 77\,nm, \tau_{0}\approx 1\,ns$) we find
$\chi_{2}\approx 10\times 4\pi$.

This decay law is assumed in the analysis of the Lancaster
experiments, in which the density of vortices is inferred from the
intensity of the signal of scattered second sound. RG improvement
leaves $^{3}He$ unchanged, but for $^{4}He$ it redefines
$\chi_{2}$ to $\chi_{2}(1-T/T_{c})^{-1/3}
> \chi_{2}$.

The first problem for the Lancaster experiments is that this makes
an already large $\chi_2$ even larger. In the attenuation of
second sound the signal to noise ratio is approximately
$O(1/\chi_{2}t)$. The empirical value of $\chi_2$ used in the
Lancaster experiments is not taken from quenches, but turbulent
flow experiments. It is suggested\cite{lancaster2} that $\chi_{2}
\approx 0.005$, a good three orders of magnitude smaller than our
prediction above. Although the TDLG theory is not very reliable
for $^{4}He$, if our estimate is sensible it does imply that
vortices produced in a {\it temperature} quench decay much faster
than those produced in turbulence.
$^{3}He$ experiments provide no check.

This is only one of our worries. We shall argue that, for early
time at least, thermal fluctuations are large in the Lancaster
experiments. However, for $^{3}He$, with negligable UV
contributions, we estimate the primordial density of
proto-vortices as
\begin{equation}
n_{zero}(t^{*})\approx \frac{{\bar t}}{8\pi (t^{*} - t_{0})
}\frac{1}{\xi_{0}^{2}}\sqrt{\frac{\tau_{0}}{\tau_{Q}}},
\label{primdef}
\end{equation}
in accord with the original prediction of Zurek. Because of the
rapid growth of $G(0,t)$, $(t^{*}-t_{0})/{\bar t} =p > 1 = O(1)$.
With $p$ behaving as $(\ln (1/{\bar\beta}))^{1/2}$ there is very
little variation. For $^{3}He$ quenches $p\approx 5$ (and for
$^{4}He$ quenches $p\approx 3$).
 We note that the factor\footnote{An errant factor of 3 appeared
 in the result of
\cite{ray}} of $f^{2}=8\pi p$ gives a value of $f = O(10)$, in
agreement with the empirical results of \cite{grenoble} and the
numerical results of \cite{zurek3}\footnote{The temperature quench
of the latter is somewhat different from that considered here, but
should still give the same results in this case}.

\subsection{Path integrals}
 Finally, we
note that $G(r, t)$, and hence its derivatives, can be expressed
as path integrals for the diffusion of a particle in a
time-dependent potential $\epsilon_{eff}$.  Thus, for example,
\begin{equation}
-G''(0, t) = \frac{T}{2}\int_{0}^{\infty} \frac{d\tau}{\tau} \oint
{\cal D}{\bf x}\,\exp\bigg[-\int_{0}^{\tau} ds\,\,
\frac{1}{4}(\frac{d{\bf x}}{d\tau})^{2} + \epsilon_{eff} (t-
s/2)\bigg]. \label{lgcorr4}
\end{equation}
In Eq.\ref{lgcorr4} the summation is over closed paths for the
particle (mass 2, in our units) traversed in time $\tau$. We note
that, if $\epsilon =\epsilon_{0}$ is fixed, then
\begin{equation}
-G''(0, t) = \frac{T}{2}\int_{0}^{\infty} \frac{d\tau}{\tau} \oint
{\cal D}{\bf x}\,e^{-L(\tau )\epsilon_{0}} \label{lgcorr5}
\end{equation}
where $L(\tau ) =\int_{0}^{\tau}d\tau '\,\sqrt{\dot{{\bf
x}}^{2}(\tau ')}$ is the length of the loop. Eq.\ref{lgcorr5} is
just the free energy of a 'free gas' of fluctuating relativistic
Brownian loops of varying lengths and shapes with tension
$\epsilon_{0}$. In conventional statistical field theory the
vanishing of $\epsilon_{0}$ in Eq.\ref{lgcorr5} signals a phase
transition due to the proliferation of loops.  The situation here
is somewhat different in that the loops have a tension that varies
with both the external time $t$ and $\tau$. The onset of the
transition is signaled by this tension vanishing at some point on
the loops.

\section{For Slow Pressure Quenches or a Large Ginzburg Regime Thermal
Fluctuations Cannot be Ignored}

\subsection{$^4 He$ experiments}

The situation for the Lancaster $^{4}He$ experiments is complex,
since they are {\it pressure} quenches for which the temperature
$T$ is almost {\it constant} at $T\approx T_{c}$.
 Unlike
temperature quenches\cite{zurek2,boyanovsky2}, thermal
fluctuations here remain at full strength\footnote{Even for $^3
He$, $T/T_{c}$ never gets very small, and henceforth we take
$T=T_{c}$ in $G_{l}(0,t)$ above}. The necessary time-{\it
independence} of $ G^{IR}(t)$ for $t>t^*$ is achieved by taking
$\epsilon_{eff} (u)= O(u^{-1})$. In consequence, as $t$ increases
beyond $t^{*}$ the relative magnitude of the UV and IR
contributions to $G_{l}(0,t)$ remains {\it approximately
constant}. Further, since for $t = t^*$,
\begin{equation}
e^{2\int_{t_{0}}^{t}du\,|\epsilon_{eff} (u)|}\,e^{-(\Delta
t)^{2}/{\bar t}^{2}}\approx 1,
\end{equation}
this ratio is the ratio at $t = t^*$.

Nonetheless, as long as the UV fluctuations are insignificant at
$t=t^*$ the density of line zeroes will remain largely independent
of scale. This follows if $ G\prime\prime^{IR}(t^*)\gg
G\prime\prime_{l}^{UV}(t^*)$, since $G\prime\prime_{l}(0,t)$
becomes scale-independent later than $G_{l}(0,t)$. In \cite{ray}
we showed that this is true provided
\begin{equation}
(\tau_{Q}/\tau_{0})(1-T_{G}/T_{c})<C\pi^{4}, \label{ginlim}
\end{equation}
where $C= O(1)$ and $T_G$ is the Ginzburg temperature. With
$\tau_{Q}/\tau_{0} = O(10^{3})$ and $(1-T_G /T_c ) = O(10^{-12})$
this inequality is well satisfied for a linearised TDLG theory for
$^{3}He$ derived\footnote{Ignoring the position-dependent
temperature of \cite{kopnin}} from the full TDGL
theory\cite{Bunkov}, but there is no way that it can be satisfied
for $^{4}He$, when subjected to a slow mechanical quench, as in
the Lancaster experiment, for which $\tau_{Q}/\tau_{0} =
O(10^{10})$, since the Ginzburg regime is so large that $(1-T_G
/T_c ) = O(1)$. Effectively, the $^{4}He$ quench is {\it nineteen}
orders of magnitude slower than its $^{3}He$ counterpart. It is
satisfying to find the Ginzburg regime reappear as an essential
ingredient in undermining universality.  In particular, as in the
original proposal of Kibble\cite{kibble1}, large thermal noise
inhibits vortex production.

When the inequality is badly violated, as with $^{4}He$ for slow
pressure  quenches, then the density of zeroes $n_{zero}=
O(l^{-2})$ after $t^*$ depends exactly on the scale $l$ at which
we look and they are not candidates for vortices. Since the whole
of the quench takes place within the Ginzburg regime this is not
implausible. However, it is possible that, even though the thermal
noise never switches off, there is no more than a postponement of
vortex production, since our approximations must break down at
some stage. The best outcome is to assume that the effect of the
thermal fluctuations on fractal behaviour is diminished, only
leading to a delay in the time at which vortices finally appear.
Even if we suppose that $n_{zero}$ above is a starting point for
calculating the density at later times, albeit with a different
$t_{0}$, thereby preserving Vinen's law, we then have the earlier
problem of the large $\chi_{2} = O(f^2 )$. In the absence of any
mechanism to reduce its value drastically, this would make it
impossible to see vortices. As a separate observation, we note
that the large value of $f^2$ in the prefactor of $n_{zero}$ is,
in itself, almost enough to make it impossible to see vortices in
$^{4}He$ experiments, should they be present. This will be pursued
elsewhere.

In summary, this work suggests that for slow pressure quenches in
$^{4}He$, we see no well-defined vortices at early times because
of thermal fluctuations, and it is plausible that, if we do see
them at later times, there are less than we would have expected
because of their rapid decay and their initial low density. The
situation is different for $^3 He$. That the density of such
vortices as appear may agree with the Zurek prediction is
essentially a consequence of dimensional analysis, given that the
main effects of the self-interaction have a tendency to cancel in
the counting of vortices, without introducing new scales. However,
a numerical simulation that goes beyond the Gaussian approximation
that is specifically tailored to the Lancaster parameters is
crucial if we are to understand the late-time behaviour and see if
this suggestion can be sustained. We shall pursue this elsewhere.

\subsection{TDLG: annular physics}

We saw for vortex formation in $^{4}He$ that the dependence of the
density on scale makes the interpretation of observations
problematic. This is not the same if the $^{4}He$ is confined to
an annulus, since the annulus itself provides the coarse-graining.
That the incoherent $\xi_{s}$ depends on its radius $l$ is
immaterial. The end result is that
\begin{equation}
\Delta v = \frac{\hbar}{m}\sqrt{\frac{1}{C\xi_{s}(t^{*};l)}}.
\label{v4}
\end{equation}
The time $t^*$  for $^{4}He$ at which we evaluate $v$, when
$\langle |\phi |^{2}\rangle = \alpha_{0}/\beta$, depends weakly on
$l$, varying from about $3{\bar t}$ to $4{\bar t}$ as $l$ varies
from $l=\xi_{0}\ll{\bar\xi}$ to $l = 10{\bar\xi}$. For $l\geq
4{\bar\xi}$ the scale at which the coarse-grained field begins to
occupy the ground states becomes largely irrelevant.

 On
expanding $f'^{2}(r,t;l)/(1 - f^{2}(r,t;l))$ in powers of $r$ and
keeping only the forward peak\cite{edik},

\begin{equation}
\frac{1}{\xi_{s}(t;l)}\approx\frac{4G_{2}} {9G_{1}}\bigg(
\frac{3G_{3}}{20G_{2}}-\frac{G_{2}}{12G_{1}}\bigg)^{-1/2},
\label{xis}
\end{equation}
where $G_{n}(r,t)$ is the $(2n-2)$th derivative of $G$ with
respect to $r$ at $r=0$ (and hence proportional to the $2n$th
moment of $G(k,t)$).

If, as suggested by Zurek, we take  $l =O({\bar \xi})$ we recover
Eq.\ref{v1} qualitatively, although a wider bore would give a
correspondingly smaller flow.

We assume\cite{edik} that
\begin{equation}
2l\geq\xi_{eff}(t;l)=\bigg(
\frac{3G_{3}}{20G_{2}}-\frac{G_{2}}{12G_{1}}\bigg)^{-1/2},
\end{equation}
since otherwise the correlations in the bulk fluid from which we
want to extract annular behaviour are of longer range than the
annulus thickness. The effect of this is to {\it reduce} the flow
velocity for narrower annuli. The effect is largest for small
radii $l\leq {\bar\xi}$, for which the approximation of trying to
read the behaviour of annular flow from bulk behaviour is most
suspect.

Once $l$ is very large, so that the power in the fluctuations is
distributed strongly across all wavelengths we recover our earlier
result, that $\xi_{s}(t^{*};l) = O(l)$. However, the change is
sufficiently slow that annuli, significantly wider than
${\bar\xi}$, for which experiments are more accessible, will give
almost the same flow as narrower annuli.  This would seem to
extend the original Zurek prediction of Eq.\ref{v1} to thicker
annuli, despite our expectations for incoherent flow.  However, we
stress again that caution is necessary, since in the approximation
to characterise an annulus by a coarse-grained ring without
boundaries we have ignored effects in the direction perpendicular
to the annulus. In particular, the circular cross-section of the
tube has not been taken into account.  One consequence of this is
that infinite (non-self-intersecting) vortices in the bulk fluid
have no counterpart in an annulus.

Since $\Delta v$ only depends on $\xi_{s}^{-1/2}$ it is not
sensitive to choice of $l > 2{\bar\xi}$ at the relevant $t$. Given
all these approximations our final estimate\cite{edik} is (in the
cm/sec units of Zurek\cite{zurek1})
\begin{equation}
\Delta v\approx 0.2(\tau_{Q}[\mu s])^{-\nu /4}/\sqrt{C[cm]}
\label{vf}
\end{equation}
for radii of $2{\bar\xi} - 4{\bar\xi}$, $\tau_{Q}$ of the order of
milliseconds and $C$ of the order of centimetres. $\nu = 1/2$ is
the mean-field critical exponent above.  In principle $\nu$ should
be renormalised to $\nu = 2/3$, but the difference to $\Delta v$
is sufficiently small that we shall not bother. Given the
uncertainties in its derivation the result Eq.\ref{vf} is
indistinguishable  from Zurek's\cite{zurek1} (with prefactor
$0.4$), but for the possibility of using somewhat larger annuli.
The agreement is, ultimately, one of dimensional analysis, but the
coefficient of unity could not have been anticipated easily, given
the small prefactors of Eq.6.20. How experiments can be performed,
even with wider annuli, is another matter.

\section{QFT: The Appearance of Structure}

It is because the formation of defects is an early-time occurrence
that it is, in part, amenable to analytic solution.  Again we
revert to the mode decomposition of Eq.\ref{mode}.  The field
becomes ordered, as before, because of the exponential growth of
long-wavelength modes, which stop growing once the field has
sampled the groundstates.  What matters is the relative weight of
these modes (the 'Bragg' peak) to the fluctuating short wavelength
modes, since the contribution of these latter is very sensitive to
the cutoff $l$ at which we look for defects. Only if their
contribution to Eq.\ref{ndef} is small when field growth stops can
a network of vortices be well-defined at early times, let alone
have the predicted density. Since the peak is non-perturbatively
large this requires small coupling, which we assume. However,
there is a problem in that, in the absence of explicit damping of
the type seen in FRW universes, rescattering of modes can rapidly
undo the early-time appearance of structure. Thus,while we take
small coupling, we do not take very small couplings of the
magnitude (e.g. $10^{-12}$) associated with inflationary models.

\subsection{ Mode Growth v Fluctuations: The free roll (continued)}

We begin by extending the analysis of Section 4.2 to later times,
still in the approximation of a {\it free} roll. This needs care
for slow quenches since the backreaction serves to hold the field
in the vicinity of the intermediate groundstates $|\phi^{2}| =
\phi_{0}^{2}(t)$ where, now
\begin{equation}
\phi_{0}^{2}(t) = \frac{-m^{2}(t)}{\lambda} =
\frac{M^{2}}{\lambda}\frac{\Delta t}{\tau_Q},
\end{equation}
where $\Delta t = t -t_0 = t - \epsilon_0\tau_Q$ as before.
Nonetheless, the free roll provides a basis for the more realistic
picture.

Prior to the completion of the quench at $\Delta t = \tau_Q$, the
mode equation (\ref{mode}), now of the form
\begin{equation}
\Biggl [ \frac{d^2}{dt^2} + {\bf k}^2  -\frac{M^{2}\Delta
t}{\tau_Q} \Biggr ]\chi^{\pm}_{k}(t)  =0, \label{mode1s}
\end{equation}
is exactly solvable, as we saw earlier.

For this section it is convenient to redefine the origin of time
at $t=t_0$, whereby we can drop the prefix $\Delta$.

 We are
primarily interested in the exponentially growing modes that
appear when
\begin{equation}
\Omega_{k}^{2}(t) = -{\bf k}^2  +\frac{M^{2}t}{\tau_Q}>0.
\end{equation}
For fixed $k$ this occurs when $t>t^{-}_{k} = \tau_Qk^{2}/M^{2}$.

The WKB solution is adequate for our purposes. The coarse-grained
$G_{l}(r;t)$ can be written as $G_{l}(r;t) =
G^{exp}(r;t)+G_{l}^{osc}(r;t)$ where
\begin{equation}
G^{exp}(r;t)\simeq \frac{T_0}{M^{2}}\int_{|{\bf k}|<k_t} d \! \!
\! / ^3 k\frac{MS_{k}(t)}{|k_{t}^{2}-k^{2}|^{1/2}} \,e^{i {\bf k}
. {\bf x}}|\alpha^{+}_{k}I_{1/3}(S_{k}(t))+\alpha^{-}_{k}
I_{-1/3}(S_{k}(t))|^{2} \label{WI}
\end{equation}
has exponentially growing long wavelength modes and
\begin{equation}
G_{l}^{osc}(r;t)\simeq \frac{T_0}{M^{2}}\int_{\Lambda >|{\bf
k}|>k_t} d \! \! \! / ^3
k\frac{MS_{k}(t)}{|k_{t}^{2}-k^{2}|^{1/2}} \,e^{i {\bf k} . {\bf
x}
}|\alpha^{+}_{k}J_{1/3}(S_{k}(t))-\alpha^{-}_{k}J_{-1/3}(S_{k}(t))|^{2}
\label{WJ}
\end{equation}
has short wavelength oscillatory modes. In both cases
\begin{equation}
S_{k}(t) = \int^{t}_{t^{-}_k}dt'\,|\Omega_{k}(t')|
=\frac{2}{3}\frac{M}{\sqrt{\tau_Q}}|t-t^{-}_k |^{3/2}.
 \label{Sk}
\end{equation}

 As in the
case of condensed matter previously, we have coarse-grained the
field by introducing a simple cut-off at $k = \Lambda = O(M)$, or
$l = \Lambda^{-1}$. For fixed $t$ the dividing monentum is
$k_{t}^{2} = M^{2}t/\tau_Q$. Provided we are far from the
transition we have incorporated the initial data into the
$\alpha^{\pm}$ in (\ref{WI}) and (\ref{WJ}). The normalisation
factor $T_{0}/M^{2}$ has been made visible. The remaining
$\alpha^{\pm}$ have no $\lambda$ dependence. Since $G^{osc}(0;t) =
O(T_{0}M)$ for $-t = O(\tau_Q )$ the $\alpha^{\pm}$ have no
$\tau_Q$ dependence.

For large $t$ the integrand in (\ref{WI}) will be peaked at some
$k_{0}(t)\rightarrow 0$ as $t\rightarrow \infty$, once the angular
integrals have been performed. Assuming that $k_{0}(t)\ll k_t$ the
upper bound in the integral can be dropped and $|k_{t}^{2}-k^{2}|$
approximated by $|k_{t}^{2}|$, knowing that there is no
singularity at $k=k_t$. With nothing to stop $|\alpha^{+}_{k}+
\alpha^{-}_{k}|^{2}$ behaving like a nonzero constant in the
vicinity of $k=0$, it can be treated as slowly varying and the
integral approximated as\cite{karra}
\begin{eqnarray}
G^{exp}(r;t)&\propto&
\frac{T}{M^{2}}\bigg(\frac{\tau_Q}{t}\bigg)^{1/2}
e^{4Mt^{3/2}/3\sqrt{\tau_Q}}\int_{|{\bf k}|<M} d \! \! \! / ^3 k
\, e^{i {\bf k} . {\bf x} } \;e^{-2\sqrt{t\tau_Q}k^{2}/M}
\nonumber
\\
&\propto&
\frac{T}{M|m(t)|}\bigg(\frac{M}{\sqrt{t\tau_Q}}\bigg)^{3/2}e^{4Mt^{3/2}/3\sqrt{\tau_Q}}
e^{-r^{2}/\xi^{2}(t)} \label{Wexp}
\end{eqnarray}
on performing the $k^{2}$ expansion of the exponent, where
\begin{equation}
\xi^{2}(t) = \frac{2\sqrt{t\tau_Q}}{M}. \label{chis}
\end{equation}
Although, like Eq.\ref{Wexp}, the expression Eq.\ref{chis} is not
supposed to be valid for small $t$, it does embody the Kibble
freezeout condition Eq.\ref{xiC} in satisfying ${\dot\xi}({\bar
t}) = O(1)$.

\subsection{Comparison with Kibble's results: First guess}

The calculations we did on CM systems showed that, although the
freeze-in time ${\bar t}$ was not relevant for the appearance of
stable vortices, their subsequent density is determined by the
scales at this time in the absence of strong fluctuations.  We
shall see that the same could be true here.

The calculations above were for a free roll. Let us suppose,
provisionally, that the backreaction exerts its influence over
such a short time that, in effect, it is if it were an {\it
instantaneous} brake to domain growth. The provisional freeze-in
time $t^{*}$ is then, effectively, the time it takes to reach the
transient groundstate $\phi_{0}^{2}(t)= -m^{2}(t)/\lambda$. That
is, $G(0;t^{*}) = O(\phi_{0}^{2}(t^{*}))$, giving
\begin{equation}
(\sqrt{t^{*}\tau_Q}M)^{3/2}e^{-4M(t^{*})^{3/2}/3\sqrt{\tau_Q}} =
O\bigg(\lambda^{1/2}(\frac{\tau_Q}{t^{*}})^{3/2}\bigg).
\label{tsff}
\end{equation}
 Neglecting $ln ln$ terms gives
\begin{equation}
Mt^{*} \simeq \frac{1}{2}(M\tau_Q)^{1/3}\bigg(\ln (1/\lambda )+
\ln (M\tau_Q )\bigg)^{2/3}. \label{tfs}
\end{equation}
The second term in the bracket of (\ref{tfs}) can be ignored to a
good approximation provided $M\tau_Q < (1/\lambda)$, preferably by
a large margin, and we assume that this is so. In terms of the
causal time ${\bar t}$ the relation is just
\begin{equation}
Mt^{*} \simeq M{\bar t}(\ln (1/\lambda ))^{2/3}.
\end{equation}
That is, the freeze in time $t^{*}$ is (qualitatively) larger than
the causal time ${\bar t}$. As far as the separation of scales is
concerned, we have the same effect qualitatively if we had taken
$t^{*}$ as the time for the field to reach the final ground state
as $|\phi^{2}| = M^{2}/\lambda$, rather than the provisional
ground states $\phi_{0}^{2}(t)$. Nonetheless, we shall remain with
(\ref{tsff}).

Whatever, the necessary condition that $t^{*} <\tau_Q $ (since the
previous calculations are for $t<\tau_Q$) reduces to
\begin{equation}
M\tau_Q > \ln (1/\lambda). \label{tauln}
\end{equation}
where, and henceforth, we neglect coefficients of $O(1)$. That is,
the quench time should be longer than the freeze-in time for the
instantaneous quench, the time it takes the field to sample the
ground-state in a free-roll\footnote{In general, we recover the
results of the instantaneous quench if we set $M\tau_Q =O(\ln
(1/\lambda))$}, but shorter than the equilibriation time (or
powers of it that preserve the separation of scales).

At this qualitative level the correlation length at the spinodal
time is
\begin{equation}
M^{2}\xi^{2}(t^{*})\simeq (M\tau_Q)^{2/3}(\ln (1/\lambda ))^{1/3}.
\label{chiss2}
\end{equation}
The effect of the other modes is larger than for the instantaneous
quench,  giving, at $t=t^{*}$
\begin{equation}
n_{zero}=  \frac{M^{2}}{\pi (M\tau_{Q})^{2/3}} (\ln (1/\lambda
))^{-1/3}[1 + E]. \label{nisMf}
\end{equation}
The error term $E =O(\lambda^{1/2}(M\tau_Q)^{4/3}(\ln
(1/\lambda))^{-1/3})$ is due to oscillatory modes, sensitive to
the cut-off. In mimicry of Eq.\ref{ndef} it is helpful to rewrite
Eq.\ref{nisMf} as
\begin{equation}
n_{zero}=  \bigg[\frac{1}{\pi
\xi_{0}^{2}}\bigg(\frac{\tau_{0}}{\tau_{Q}}\bigg)^{2/3}\bigg] (\ln
(1/\lambda ))^{-1/3}[1 + E]. \label{nisMf2}
\end{equation}
in terms of the scales $\tau_{0} = \xi_{0} = M^{-1}$. The first
term in Eq.\ref{nisMf2} is the Kibble estimate of Eq.\ref{ndef},
the second is the multiplying factor, rather like that in
Eq.\ref{primdef}, that yet again shows that estimate can be
correct, but for completely different reasons. As for condensed
matter, the dependence on the interaction strength is only through
a power of the logarithm of $(1/\lambda ))^{-1/3}$.

The third term shows when it can be correct, since $E$ is also a
measure of the sensitivity of $n_{zero}$ to the scale at which it
is measured. The condition $E^{2}\ll 1$, necessary for a vortex
network to be defined, is then guaranteed\cite{ray} if
\begin{equation}
(\tau_{Q}/\tau_{0})^{2}(1-T_{G}/T_c)<C, \label{ginlim2}
\end{equation}
where $C = O(1)$, on using the relation $(1-T_{G}/T_c) = O(\lambda
)$. This is the QFT counterpart to Eq.\ref{ginlim} and concurs
again, in principle, with Kibble's earlier argument that large
thermal fluctuations inhibit the appearance of vortices.

The easiest way to enforce $E\ll 1$ and  $M\tau_{Q} > \ln
(1/\lambda )$ is to take $M\tau_{Q} = \ln (1/\lambda )^{\alpha}$,
for $\alpha
>1$. The effect in Eq.\ref{nisMf2} is merely to renormalise the
critical index.  Of course, the Kibble prediction Eq.\ref{ndef}
was only an estimate.  Although it is good qualitatively it
 is  misleading  when considering definition of the
network since a simple calculations shows that, at time ${\bar
t}$, the string density is still totally sensitive to the
definition of coarse-graining.

Finally, suppose that this approach is relevant to the local
strings of a strong Type-II $U(1)$ theory for the early universe,
in which the time-temperature relationship $tT^{2} = \Gamma
M_{pl}$ is valid, where we take $\Gamma = O(10^{-1})$ in the GUT
era.  If $G$ is Newton's constant and $\mu$ the classical string
tension then, following \cite{zurek1}, $M\tau_Q\sim
10^{-1}\lambda^{1/2}(G\mu)^{-1/2}$.  The dimensionless quantity
$G\mu\sim 10^{-6} - 10^{-7}$ is the small parameter of cosmic
string theory.  A value $\lambda\sim 10^{-2}$ gives $M\tau_Q\sim
(Mt^{*})^{a}, a\sim 2$, once factors of $\pi$, etc.are taken into
account, rather than $M\tau_Q\sim 1/\lambda$, and the density of
Eq.\ref{nisMf2} may be relevant.

\subsection{Backreaction in QFT}

To improve upon the free-roll result more honestly, but retain the
Gaussian approximation for the field correlation functions, the
best we can do is adopt a mean-field approximation along the lines
of \cite{boyanovsky,LA}, as we did for the CM systems earlier. As
there,  it does have the correct behaviour of stopping domain
growth as the field spreads to the potential minima.  As before,
only the large-$N$ expansion preserves Goldstone's theorem.

$G(r;t)$ still has the mode decomposition of (\ref{modesum}), but
the modes $\chi^{\pm}_{k}$ now satisfy the equation
\begin{equation}
\Biggl [ \frac{d^2}{dt^2} + {\bf k}^2 + m^2(t) +
\lambda\langle\Phi^{2}({\bf 0})\rangle_{t} \Biggr
]\chi^{\pm}_{k}(t) =0, \label{modeh}
\end{equation}
where we have taken $N=2$. Because $\lambda\phi^{4}$ theory is not
asymptotically free, particularly in the Hartree approximation,
the renormalised $\lambda$ coupling shows a Landau ghost. This
means that the theory can only be taken as a low energy effective
theory.

The end result is\cite{boyanovsky}, on making a single subtraction
at $t=0$, is
\begin{equation}
\Biggl [ \frac{d^2}{dt^2} + {\bf k}^2 + m^2(t) + \lambda\int d \!
\! \! / ^3 p \, C(p) [\chi^{+}_{p}(t)\chi^{-}_{p}(t)-1] \Biggr
]\chi^{\pm}_{k}(t) =0. \label{modeh2}
\end{equation}
which we write as
\begin{equation}
\Biggl [ \frac{d^2}{dt^2} + {\bf k}^2 -\mu^2(t) \Biggr
]\chi_{k}(t)  =0. \label{modemu}
\end{equation}
On keeping just the unstable modes in $\langle\Phi^{2}({\bf
0})\rangle_{t}$ then, as it grows, its contribution to
(\ref{modeh2}) weakens the instabilities, so that only longer
wavelengths become unstable. At $t^{*}$ the instabilities shut
off, by definition, and oscillatory behaviour ensues. Since the
mode with wavenumber $k >0$ stops growing at time $t^{+}_k
<t^{*}$, where $\mu^{2}(t^{+}_{k}) = {\bf k}^{2}$, the free-roll
density at $t^{*}$ must be an overestimate.

An approximation that improves upon the WKB
approximation\cite{karra} is
\begin{equation}
\chi_{k}(t) \approx \bigg(\frac{\pi M}{2\Omega_{k}(\eta
)}\bigg)^{1/2} \exp\bigg(\int_{0}^{t}dt\,\Omega (t)\bigg)
\label{chiKf}
\end{equation}
when $\eta =M(t^{+}_k -t) > 0$ is large, and $\Omega_{k}^{2}(t) =
\mu^{2}(t) - {\bf k}^{2}$. On expanding the exponent in powers of
$k$ and retaining only the quadratic terms we recover the WKB
approximation when $\mu (t)$ is non-zero.

The  result is that the effect  of the back-reaction is to give a
time-delay $\Delta t$ to $t^*$, corresponding to a decrease in the
value $k_{0}(t)$ at which the power peaks of order
\begin{equation}
\frac{\Delta t}{t^*} = O\bigg(\frac{1}{ln(1/\lambda)}\bigg).
\label{lag}
\end{equation}
The backreaction has little effect for times $t<t^{*}$. For
$t>t^{*}$ oscillatory modes take over the correlation function and
we expect oscillations in $G(k;t)$.

In practice the backreaction rapidly forces $\mu^{2}(t)$ towards
zero if the coupling is not too small\cite{boyanovsky}. For
couplings that are not too weak the end result is a new power
spectrum, obtained by superimposing oscillatory behaviour onto the
old spectrum. As a  gross oversimplification, the contribution
from the earlier exponential modes alone can only be to contribute
terms something like
\begin{eqnarray}
G(r;t)&\propto&
\frac{T}{M|m(t^{*})|}e^{4M(t^{*})^{3/2}/3\sqrt{\tau_Q}}
\int_{|{\bf k}|<M} d \! \! \! / ^3 k \, e^{i {\bf k} . {\bf x} }
\;e^{-2\sqrt{t^{*}\tau_Q}k^{2}/M} \nonumber
\\
&\times&\bigg[\cos k(t-t^{*}) +\frac{\Omega (k)- W'(k)}{k}\sin
k(t-t^{*})\bigg]^{2} \label{Wexps}
\end{eqnarray}
to $G$, where $\Omega = M(t^{*} - t_k)^{1/2}/\tau_Q^{1/2}$ and $W'
= 1/4(t^{*} -t_k )$.  Fortunately, the details are almost
irrelevant, since the density of line zeroes is independent of the
normalisation, and only weakly dependent on the power spectrum.

The $k=0$ mode of Eq.\ref{Wexps} encodes the simple solution
$\chi_{k=0}(t) = a + bt$ when $\mu^2 = 0$. As
observed\cite{boyanovsky2} by Boyanovsky et al. this has built
into it the basic causality discussed by Kibble\cite{kibble3}.
Specifically, for $r,t\rightarrow\infty$, but $r/t$ constant
$(\neq 2)$,
\begin{equation}
G(r,t)\approx \frac{C}{r}\Theta (2t/r -1).
\end{equation}
It has to be said that this approximation should not be taken very
seriously for large $t$, since we would expect rescattering to
take place at times $\Delta t = O(1/\lambda)$ in a way that the
Gaussian approximation precludes.

Whatever, it follows directly that this causality, engendered by
the Goldstone particles of the self-consistent theory, has little
effect on the density of line-zeroes that we expect to mature into
fully classical vortices. If Eq.\ref{ginlim2} is not satisfied, it
is difficult to imagine how clean vortices, or proto-vortices, can
appear later without some additional ingredient.

\subsection{Really slow quenches}

Finally, consider slowing the quench until the WKB approximation
manifestly breaks down. This is almost certainly the case if the
growing modes catch up with the moving minima within the Ginzburg
regime by time
\begin{equation}
t_G = O(\lambda \tau_Q ).
\end{equation}
If we take (\ref{tsff}) seriously  it follows that domain growth
has stopped by time $t_G$ provided $M\tau_Q$ is a power of
$\lambda^{-1}$. Equation (\ref{tsff}) only makes sense if $Mt_G\gg
1$ i.e  $M\tau_Q\gg \lambda^{-1}$. As an example, let us take
$M\tau_Q = \lambda^{-3/2}$. In this case the dynamical correlation
length $\xi (t_G )$ of (\ref{chis}) is $O(1/M\lambda^{1/2})$,
equal to the equilibrium correlation length $\xi_{eq}(t_G ) =
|m^{-1} (t_G )|$,  suggesting the string density $n_{G} =
O(M^{2}/\lambda )$ that follows from (\ref{tom1}), very much
smaller than that of (\ref{nisMf2}) above. However, a simple check
shows that, in this case,
\begin{equation}
\frac{l}{n_{l}(t_{G})}\frac{\partial n_{l} (t_G)}{\partial l
}\bigg|_{l = M^{-1}} = O(1).
\end{equation}
Despite the low density of line zeroes at the Ginzsburg
temperature, they do not provide a  stable network of
proto-vortices.

\section{Conclusions}

We examined the Kibble /Zurek predictions for the onset of phase
transitions and the appearance of defects (in particular, vortices
or global cosmic strings) as a signal of the symmetry breaking.
Our results are in agreement with their prediction Eq.\ref{xiKZ}
as to the magnitude of the correlation length at the time the
transition truly begins, equally true for condensed matter and
QFT.

However, this is not simply a measure of the separation of defects
at the time of their appearance.  The time ${\bar t}$ is too early
for the field to have found the true groundstates of the theory.
We believe that time, essentially the spinodal time, is the time
at which proto-vortices can appear, which can later evolve into
the standard classical vortices of the theory.

Even then, they may be frustrated by thermal field fluctuations.
In TDLG condensed matter thermal noise is proportional to
temperature. If temperature is {\it fixed}, but {\it not}
otherwise, as in the pressure quenches of $^{4}He$ this noise can
inhibit the production of vortices, although there are other
factors to be taken into account (such as their decay rate). On
quenching from a high temperature in QFT there are always thermal
fluctuations, and these can also disturb the appearance of
vortices.
 The condition that thermal fluctuations are ignorable at the time
that the field has achieved the true ground-states can be written
\begin{equation}
(\tau_{Q}/\tau_{0})^{\gamma}(1-T_{G}/T_c )<C, \label{ginlim3}
\end{equation}
where $\gamma = 1$ for condensed matter and $\gamma = 2$ for QFT.
$C = O(1)$.

This restores the role of the Ginzburg temperature $T_G$ that the
simple causal arguments overlooked. As in the earlier arguments,
large fluctuations inhibit vortex production, but in their absence
the equilibrium correlation length is not the relevant quantity.
Quenches in $^{4}He$ provide the major example for which
Eq.\ref{ginlim3} is not satisfied. However, when Eq.\ref{ginlim3}
is satisfied, the Kibble/Zurek prediction is recovered for the
density of line zeroes, now seen as potential vortices. That no
new scales (up to logarithms) are introduced by the interaction is
a reflection of the fact that line-zero density only uses limited
properties of the power spectrum of fluctuations (in the Gaussian
approximation at least).

What happens at late time is unclear, although for TDLG numerical
simulations can be performed (but have yet to address this problem
exactly). On the other hand, not only is the case of a single
self-interacting {\it quantum} scalar field in flat space-time a
caricature  of the early universe, but it is extremely difficult
to go beyond the Gaussian approximation.  To do better requires
that we do differently. There are several possible approaches. One
step is to take the FRW metric of the early universe seriously,
whereby the dissipation due to the expansion of the universe can
change the situation dramatically\cite{stephens}. Other approaches
are more explicit in their attempts to trigger decoherence
explicitly, as we mentioned earlier. Most simply, one treats the
short wavelength parts of  the field as an environment to be
integrated over, to give a coarse-grained theory of
long-wavelength modes acting classically in the presence of noise.
However, such noise is more complicated than in TDLG theory, being
multiplicative as well as additive, and
coloured\cite{gleiser2,lombardo,muller2}. This is an area under
active consideration, but we shall stop here.

I would like to thank Alasdair Gill, Tom Kibble, Wojciech Zurek
and Grisha Volovik for fruitful discussions. There is much work in
this area and we have not included all references.  We apologise
to any authors who we may have missed inadvertently. This work is
the result of a network supported by the European Science
Foundation.

\end{document}